\documentclass[
 reprint,
onecolumn,
preprint,
 amsmath,amssymb,
 aps,
]{revtex4-1}

\usepackage{graphicx}
\usepackage{dcolumn}
\usepackage{stackrel}
\usepackage{bm}

\usepackage{enumerate}

\begin{document}


\title{Instantaneous charge state of Uranium projectiles in fully ionized plasmas from energy loss experiments}

\author{Roberto Morales}
\author{Manuel D. Barriga-Carrasco}
\email{ManuelD.Barriga@uclm.es}
\affiliation{E.T.S.I. Industriales, Universidad de Castilla-La Mancha, E-13071 Ciudad Real, Spain 
}

\author{David Casas}
 \affiliation{E.T.S.I. Industriales, Universidad de Castilla-La Mancha, E-13071 Ciudad Real, Spain}
 \affiliation{Max Born Institute, Max Born Str. 2a D-12489, Berlin, Germany 
}

\date{\today}

\begin{abstract}
The instantaneous charge state of uranium ions traveling through a fully ionized hydrogen plasma has been theoretically studied and compared with one of the first energy loss experiments in plasmas, carried out at GSI-Darmstadt by Hoffmann \textit{et al.} in the 90's. For this purpose, two different methods to estimate the instantaneous charge state of the projectile have been employed: (1) rate equations using ionization and recombination cross sections, and (2) equilibrium charge state formulas for plasmas. Also, the equilibrium charge state has been obtained using these ionization and recombination cross sections, and compared with the former equilibrium formulas. The equilibrium charge state of projectiles in plasmas is not always reached, it depends mainly on the projectile velocity and the plasma  density. Therefore, a non-equilibrium or an instantaneous description of the projectile charge is necessary. The charge state of projectile ions cannot be measured, except after exiting the target, and experimental data remain very scarce. Thus, the validity of our charge state model is checked by comparing the theoretical predictions with an energy loss experiment, as the energy loss has a generally quadratic dependence on the projectile charge state. The dielectric formalism has been used to calculate the plasma stopping power including the Brandt-Kitagawa (BK) model to describe the charge distribution of the projectile. In this charge distribution, the instantaneous number of bound electrons instead of the equilibrium number has been taken into account. Comparing our theoretical predictions with experiments, it is shown the necessity of including the instantaneous charge state and the BK charge distribution for a correct energy loss estimation. The results also show that the initial charge state has a strong influence in order to estimate the energy loss of the uranium ions. 
\end{abstract}

\pacs{Valid PACS appear here}
\maketitle

\section{\label{sec:Introduction} Introduction}
The charge state of heavy ions in cold matter (solid and gas targets) has been studied over a long period of time theoretically as well as experimentally \cite{BellPR53,Betz83}. Due to the high densities of the solid targets, an equilibrium charge state is usually reached after a short distance \cite{Bohr48,NorthcliffePR60,KreusslerPRB81,SRIM}. Many experimental data have been collected and several empirical and semi-empirical formulas have been developed to estimate the charge state of the projectile as a function of its velocity and its atomic number with a very good agreement with the data at high velocities \cite{Bohr54,Betz72,Wittkower-Betz1973,Shima89}. A few codes are also available to calculate the charge state distribution and/or ionization and recombination cross sections, which determines the projectile charge state, in the case of cold matter \cite{RozetNIMB96,LamourPRA15,ShevelkoNIMB10,LitsarevCPC13}. 

In the case of plasma targets, there are no sufficient experimental data to develop semi-empirical formulas as a fit of these experiments. Moreover, in the plasma case, there are not only bound electrons in the target, but also free electrons that modify the ionization and recombination processes and therefore the projectile charge state.

The study of the charge state of heavy ions in plasmas is important, for example, on its role for the energy loss of these projectile ions in the target, as the energy loss depends for a linear beam-plasma interaction quadratically on the charge state. The energy loss is relevant for many applications in different fields of science, from fast ignition inertial fusion to medical applications \cite{OgawaFED99,EliezerLPB07,HoffmannNIMA07,CookLPB08,BarrigaPP08}. Modeling  the charge state of heavy ions in plasmas is also relevant for new plasma stripper techniques in accelerator physics \cite{ChabotPRE95,OguriNIMB00}.

In the last three decades, a few experiments to measure the energy loss and the charge state of heavy ions traveling through ionized matter have been carried out \cite{WeyrichNIMPRA89,HoffmannPRA90,GardesPRA92,DietrichPRL92,GardesPA92,CouillaudPRE94,JacobyPRL95,KojimaLPB02,SkobelevNIMB05,FrankPRL13,GauthierPRL13}. In these experiments, two main effects of the ionized matter have been confirmed: the enhanced plasma energy transfer (EPET), which means a larger energy loss of ions in plasmas than in cold matter due to a more efficient energy transfer to the free electrons, and the enhanced projectile ionization in plasma (EPIP), which means a higher projectile charge state in plasmas than in cold matter mainly due to the reduction of the capture cross sections with target free electrons.

The charge state of ions in matter is established through a competition between electron loss (projectile ionization) and electron capture (projectile recombination) processes. The charge state of an ion beam is increased in fully ionized plasmas compared to cold matter because the cross section for the capture of a target free electron is much smaller than the one for the capture of a target bound electron \cite{BellPR53}.  The reason is that direct free electron capture by a moving projectile, due to momentum conservation, is a three-body collision process, and the probability for a free electron to find a third collision partner is smaller than for a bound one. The EPIP was first theoretically studied by Nardi and Zinamon \cite{NardiPRL82} and by Peter and Meyer-ter-Vehn \cite{PeterPRA91b} later, and it was also measured by Dietrich et al. \cite{DietrichPRL92}. 

The aim of this work is to develop two methods to estimate the instantaneous charge state of heavy ions penetrating fully ionized plasmas. The first one is a more elaborated model; the instantaneous charge state is estimated using ionization and recombination cross sections. Due to the complexity of calculating the recombination cross sections, a simple analytical model has been also developed, where the instantaneous charge state is estimated using equilibrium charge state formulas and only ionization cross sections. Then, these models can be included in an energy loss description to compare the theoretical predictions with experimental data for an uranium energy loss in a fully ionized hydrogen plasma.

In section \ref{sec_Projectile_charge_state} the two different methods are described. First, all ionization and recombination processes are formally presented by means of their cross sections. Then, the instantaneous charge state as well as the equilibrium charge state are obtained by solving the rate equations that govern these charge-exchange processes. Moreover, former equilibrium formulas for plasmas are briefly discussed in this section and the new analytical model is proposed.

In section \ref{sec_results}, a brief description of the energy loss model used in this work is presented. Here, the dielectric formalism is used to estimate the stopping power of the plasma free electrons when interacting with a projectile. Also, the equilibrium charge state as well as the instantaneous charge state, according to the two different methods, are compared with the Monte Carlo code described by Hoffmann \textit{et al.} in the analyzed experiment. Finally, the validity of the charge state models is checked by comparing the theoretical estimations with the energy loss experiment and very good agreement has been found.

The validity of this work is restricted to the ideal case of fully ionized plasmas, i.e., the ionization degree of the plasma target is supposed to be $100\%$, and for plasma free electron densities of up to $n_e \simeq 10^{20}$ cm$^{-3}$, for which the so-called density effect, which means a reduction of the recombination cross sections due to the high density, can be neglected \cite{PeterPRA91b}. 

It is also restricted to high projectile velocities, when the energy loss is supposed to be small and the projectile velocity can be assumed as constant, and to the linear regime \cite{PeterPRA91b}, 
$$
\gamma = \frac{Z}{n_e \lambda_D^3} \frac{1}{1 + v^3 / (k_B T/m_e)^{3/2}} \ll 1,
$$
and to weakly coupled and non-degenerate plasmas, 
$$
\Gamma = \frac{e^2 (4 \pi n_e / 3)^{1/3}}{k_B T} \ll 1, \qquad \theta = \frac{k_B T}{E_F} \gg 1.
$$
Here, $v$ and $Z$ are the projectile velocity and atomic number, $\lambda_D=(k_B T /4 \pi e^2 n_e )^{1/2}$ is the Debye length, $k_B$ is the Boltzmann constant, $n_e$ is the plasma free electron density, $T$ is the plasma temperature, $e$ and $m_e$ are respectively the electron charge and electron mass, $E_F = 1/2\, m_e v_F^2$, $v_F$ being the Fermi velocity and $\Gamma$ and $\theta$ are the non-ideality parameter and the degeneracy parameter, respectively.

\section{\label{sec_Projectile_charge_state} Instantaneous projectile charge}

		\subsection{Method I. Cross sections model}\label{sec_rate_equations}
The possible charge states for an ion beam are between $0$, i.e., neutral atom with all its electrons bound to the nucleus of the projectile, and $Z$, i.e., projectile fully stripped of all its electrons. The charge state is determined by a dynamical equilibrium established by all charge-exchange processes,  which are governed by projectile electron loss and capture cross sections, between $q$ and $q^\prime$, i.e.,
\begin{equation}
\frac{dF_q(t)}{dt}=\sum_{q^\prime\not= q}{\alpha (q^\prime\rightarrow q)F_{q^\prime}(t)} -  \sum_{q^\prime\not= q}{\alpha (q\rightarrow q^\prime)F_{q}(t)},
	\label{ec_conservation_of_particles_1}
\end{equation}
where $\alpha (q'\rightarrow q)$ and $\alpha (q\rightarrow q')$ are the total rate of all charge-exchange processes between $q$ and $q^\prime$ and $F_q$ is the fraction of projectiles with charge state $q$. Here,
$$
F_q, \ \ \forall \ q \in\left[0-Z\right]
$$
represent the charge state distribution of the projectile, $q$ being any integer value between $0$ and $Z$, and $\alpha_i = \sigma_i n v$, where $\sigma_i$ is the cross section of the process $i$ and $n$ is the target density. The  $\alpha_i$ rates might include the loss and capture of more than one electron simultaneously. However, the mono-electronic processes in most cases dominate the loss and capture of two or more electrons simultaneously \cite{Betz83}, and therefore, only single-loss and -capture processes have been included in this work.

These loss, $L_q$, and capture, $C_q$, mono-electronic rates can be defined as,
$$
\alpha(q\rightarrow q+1)\equiv L_q, \qquad \alpha(q\rightarrow q-1)\equiv C_q,
$$
and equation (\ref{ec_conservation_of_particles_1}) becomes,
\begin{equation}
\frac{dF_q(t)}{dt}=C_{q+1}F_{q+1}(t)+L_{q-1}F_{q-1}(t)-(C_q+L_q)F_q(t),
	\label{ec_conservation_of_particles_2}
\end{equation}
with the normalization condition $\sum_{q=0}^{Z}{F_q}(t)=1$. Then, the instantaneous charge state according to the cross sections model can be obtained as,
\begin{equation}
	Q(x)=\sum_{q=0}^{Z}{qF_q(x)},
	\label{eq_instantaneous_charge_state_cross_sections_model}
\end{equation}
where $x = v t$ is the depth (or distance) traveled by the projectile.

The equilibrium charge state can be also calculated assuming the stationary case, i.e., $dF_q(t)/dt =0$, and then the equation (\ref{ec_conservation_of_particles_2}) is reduced to \cite{Betz72},
\begin{equation}
L_qF^{eq}_q=C_{q+1}F^{eq}_{q+1},
\end{equation}
where $F_q^{eq}$ denotes the charge state distribution of the projectile once the equilibrium is reached. Then, the equilibrium charge state can be obtained as,
\begin{equation}
Q_{eq}=\sum_{q=0}^{Z}{qF^{eq}_q}.
\label{eq_eq_cross_sections_model}
\end{equation}

In the following, the models used to calculate the ionization (projectile electron loss) and recombination (projectile electron capture) rates are briefly described. The present work is based on simple atomic modeling and uses screened hydrogenic energies and oscillator strengths. In the literature, several models are available to calculate the screening coefficients for the hydrogenic model like the Faussurier model \cite{FaussurierJQSRT97}, or the Slater's rules \cite{SlaterPR30}. In this work, the Slater's rules have been used.

\subsubsection{Ionization rates, $L_q$}\label{sec_ionization_cross_sections}

A projectile penetrating a fully ionized plasma can lose electrons due to collision processes with (a) the plasma ions and (b) the plasma free electrons. \\

\paragraph{Ionization with plasma ions} 

$$X^{q+}+A^{p+}\rightarrow X^{(q+1)+}+A^{p+} + e,$$

Here, $X$ is the projectile ion with charge state before and after ionization $q+$ and $(q+1)+$ respectively, $A$ is the plasma target ion with charge state $p+$ and $e$ is the ionized electron.

One of the simplest and most robust models to quantify the ionization cross section of a projectile in its interaction with the target ions is based on the semi-classical approximation given by the binary-encounter model (BEM), used first by Gryzinski to calculate cross sections for charge-exchange and ionization processes \cite{GryzinskiPR65a, GryzinskiPR65b}. According to this model, the cross section for the ionization of an electron bound in the $n_{th}$ shell of the projectile in a collision with the target ion is given by
\begin{equation}
		\sigma_{BEM}=\sum_n{N_n\sigma_n}=\sum_n{N_n\sigma_0 \left[\frac{Z^*_{t}}{U_n}\right]^2 G \left(\frac{v}{v_n}\right)},
		\label{eq_sigma_BEM}
\end{equation}
where $\sigma_{BEM}$ is expressed in cm$^2$, $N_n$ is the number of electrons bound to the projectile shell $n$, $U_n$ is its binding energy, \mbox{$\sigma_0=6.56\times 10^{-14}$ cm$^2$eV$^2$} and $ v_n=\left(2U_n/m_e\right)^{1/2} $ is the electron orbital velocity. The function $ G (u) $ reaches its maximum when $ u = 1 $ and represents the matching condition $v\simeq v_n$ for maximum ionization. When $ u \geq 0.206 $, $ G (u) $ is given by \cite{McGuirePRA73}
\begin{eqnarray}
G(u) & = & \left[\frac{u^2}{(1+u^2)}\right]^{\frac{3}{2}}\nonumber\\ 
& & \times u^{-2}\left[ \left(\frac{u^2}{1+u^2}\right)+\frac{2}{3}\left(1+\frac{1}{\beta}\right)\ln(2.7+u)\right] \nonumber\\ 
& & \times \left[1-\frac{1}{\beta}\right]\left[1-\left(\frac{1}{\beta}\right)^{1+u^2}\right],
\end{eqnarray}
where $\beta=4u^2\left(1+1/u\right)$. When $u<0.206$, \mbox{$G(u)=4u^4/15$.}

In the case of a fully ionized plasma, $Z^*_t$ is the nuclear charge of the target ions. If the plasma is partially ionized, the nuclear charge is partially screened by the bound electrons, and an effective target charge $Z^*_t$ must be calculated. For the particular case of a fully ionized hydrogen plasma, $Z^*_t=Z_t=1$. 

Then, the ionization rate, in units $s^{-1}$, is given by
\begin{equation}
\alpha_{BEM}=\sigma_{BEM} n_i v,
\end{equation}
where $ n_i $ is the target ion density.\\

\paragraph{Ionization with plasma free electrons}

$$X^{q+}+e\rightarrow X^{(q+1)+}+2e.$$

Following Lotz, the ionization cross section in collisions with free electrons can be adapted as \cite{LotzZP67, LotzZP68},
\begin{equation}
		\sigma_{FE}=4\times 10^{-14} \sum_n N_n\frac{\ln (E_r/U_n)}{E_rU_n}\theta (E_r-U_n),
		\label{eq_sigma_IFE}
		\end{equation}
where $E_r=(m_e/2)v_r^2$ is the relative energy between the projectile and the plasma free electrons, \mbox{$v_r \simeq (v^2 + v_ {the}^2)^{1/2}$}, where $ v_ {the} = (k_BT/m_e)^{1/2}$ is the thermal velocity of the plasma free electrons. 

In equation (\ref{eq_sigma_IFE}), the step function $ \theta = (|x|+x)/2 $ marks the limit in which a free electron does not have a sufficient kinetic energy to ionize an electron of the projectile. In fact, that is the main difference between ionizing collisions with target ions and target free electrons. 

Then, the ionization rate is given by
\begin{equation}
\alpha_{FE}=\sigma_{FE} n_e v_r.
\end{equation}

Finally, the total loss rate is estimated as,
\begin{equation}
L_q=\alpha_{BEM} + \alpha_{FE},
\label{ec_total_loss_rate}
\end{equation}
evaluated for the charge state $q$.

\subsubsection{Recombination rates, $C_q$}
Due to momentum conservation, the projectile can only capture an electron in the presence of a third collision partner. For the capture of a free electron, the third partner can either be a photon [radiative electron capture (REC)], another bound electron of the projectile ion [dielectronic recombination (DR)], or a third free electron [three-body recombination (3BR)].\\

\paragraph{Radiative Electron Capture}

$$X^{q+}+e\rightarrow X^{(q-1)+}+h\nu.$$

The cross section for radiative electron capture (REC) is related to the electronic transition probability $A(n^\prime\rightarrow n)$ between major shells $n^\prime\rightarrow n$ in the projectile with charge state $q$ and the emission of a photon with energy \mbox{$h\nu=E_{n^\prime}-E_n$.}

There are several theoretical models to estimate the rate of the radiative electron capture \cite{MenzelAJ37,SpitzerAJ48,SeatonRAS59}. In our model, an analytical formula developed by Peter has been used \cite{PeterLPB90},
\begin{equation}
\alpha_{REC} = \frac{2^6}{3}\left(\frac{\pi}{3}\right)^{1/2}a_0^2 c \alpha^4 n_e q^2 \frac{0.78}{v}  \frac{x_v^{0.3}}{1+x_v^2},
\label{equation_alpha_REC}
\end{equation}
where $ a_0 $ is the Bohr radius, $ \alpha $ is the fine structure constant, $ c $ is the light velocity in vacuum and $x_v$ is given by
\begin{equation}
x_v=\frac{v}{v_n} \simeq \frac{v}{q}\left(\frac{3}{2}\left(Z-q\right)\right)^{1/3}.
\label{eq_x_Peter}
\end{equation}

\paragraph{Dielectronic Recombination}

$$X^{q+}+e\rightarrow (X^{(q-1)+})^{**}\rightarrow X^{(q-1)+} + h\nu.$$

The dielectronic recombination is a two-step process in which a free electron (kinetic energy $E_k$) is captured by a projectile ($X$, with charge state $q$) into a level $n\equiv (n_nl_n)$ and the excess energy is transferred to another electron already bound in shell $i\equiv (n_il_i)$, which is excited to shell $j\equiv (n_jl_j)$. Here, $n$ and $l$, denote the principal and the azimuthal quantum numbers, respectively. 

The free electron tends to be captured into a highly excited level, i.e., $n_n \gg 1$, from where it will usually autoionize by the process,
$$
X(q,i)+e(E_k,l\pm 1) \stackbin{A_a}{\rightarrow} X^{**}(q-1,jn)
$$
running from right to left (Auger effect with rate $A_a$). In order to stabilize $X^{**}$, the energy has to leave the system by other means, in this model via a stabilizing radiative decay (with rate $A_r$),
$$
X^{**}(q-1,jn)\stackbin{A_r}{\rightarrow} X^{*}(q-1,in) + h\nu,
$$
and the photon carries out the energy $h\nu = E_j-E_i$. Because of its longer lifetime, the higher excited electron remains in the $n\equiv (n_nl_n)$ level after the $i \rightarrow j$ transition, until it finally cascades down to the ground state. The dielectronic recombination is therefore a resonance process with the matching energy condition,
\begin{equation}
E_k+E_n=E_i-E_j.
\end{equation}

For this process, Nardi assumed a constant value of $10^{- 11}$ cm$^{3}$/s which is believed to be an upper-limit estimate, since it is the largest value quoted \cite{HahnPRA80}. Several authors have tried to model this process \cite{PeterPRL86,DasguptaPRA90,FournierPRA97}. In this work, the model described by Peter \textit{et al.} has been employed because it has been used successfully in the interpretation of a few experiments \cite{DietrichPRL92,FrankPRL13}. 

The dielectronic recombination rate is given by \cite{PeterPRL86}
\begin{eqnarray}
\alpha_{DR} & = & \frac{h^3n_e}{\left(2\pi m_e v_{the}^2\right)^{3/2}}  \sum_{n_i,l_i}\sum_{n_j,l_j}\sum_{n_n,l_n}N_{n_il_i}\frac{g_{l_j}-N_{n_jl_j}}{g_{l_j}} \nonumber \\
& & \times \left(2l_n+1\right)\frac{A_r^{(1)}A_a^{(1)}}{A_r^{(1)}+A_a^{(1)}}F(a,b),
\end{eqnarray}
where $ h $ is the Planck constant, $ N_ {n_il_i} $ and $ N_ {n_jl_j} $ are the occupation number of the $ n_il_i $ and $ n_jl_j $ shells, respectively, \mbox{$ g_ {l_j} = 2(2l_j+1)$} and,
\begin{equation}
F(a,b)=\frac{e^{-\left(a-b\right)^2}-e^{-\left(a+b\right)^2}}{4ab},
\end{equation}
with,
$$
a=\left[\frac{E_{n_jl_j}-E_{n_il_i}+E_{n_nl_n}}{k_BT}\right]^{1/2},\qquad b=\left[\frac{m_ev^2}{2k_BT}\right]^{1/2},
$$
where $ E_ {n_il_i} $, $ E_ {n_jl_j} $ and $ E_ {n_nl_n} $ are, respectively, the binding energy of the electrons in the shells $ n_il_i $, $ n_jl_j $, and $n_nl_n$. 

The rates $A_r  ^ {(1)} $ for radiative stabilization and $ A_a ^ {(1)} $ for autoionization are given by
\begin{equation}
A_r^{(1)}=\frac{1\, \text{Ry}}{\hbar}\alpha^3\left[\frac{E_{n_jl_j}-E_{n_il_i}}{1\, \text{Ry}}\right]^2f^{(1)}(j\rightarrow i),
\end{equation}
\begin{eqnarray}
A_a^{(1)} & = & \frac{8}{\sqrt{3}}\frac{1\, \text{Ry}}{\hbar}\frac{q}{n^3}\frac{1\, \text{Ry}}{E_{n_il_i}-E_{n_jl_j}}\frac{1}{2l+1}\nonumber \\
& & \times f^{(1)}(i\rightarrow j)N_0e^{-d\left(l-l_p\right)^2},
\end{eqnarray}
where $ f ^ {(1)} (i \rightarrow j) $ is the absorption oscillator strength of the $ i \rightarrow j $ excitation, approximated by the results for hydrogen \cite{QMOTEA_Bethe},  Ry$=13.6$ eV and 
\begin{equation} 
N_0 = \frac {0.4} {\sqrt {\pi}} \sqrt {d} \left[ 2- \exp{(- 0.6 / n_n ^ {4/3} \sqrt {d}})\right] ^ { -1},
\end{equation}
where $l_p=n_n^{2/3}(1 +y)^{1/2}$, $d=(1+y)^{1/2}/2.5n_n$, $y =n_n^2E_k/q^2$ Ry and $E_k=(m_e/2)v^2$.\\

\paragraph{Three-body Recombination}

$$X^{q+}+e+e\rightarrow X^{(q-1)+}+e.$$

In plasmas with high density (similar to solid state densities), the probability for a simultaneous collision of the projectile with two electrons increases. Therefore, one of these free electrons can be captured by the projectile, while the other one carries away the excess energy. The capture of free electrons by the three-body recombination process (3BR) can be calculated using the description of Zel'dovich and Raizer \cite{ZeldovichRaizer66} of the Thomson classical theory \cite{Massey52},
\begin{equation}
\alpha_{3BR}=2.92\times 10^{-31}\frac{q^3(n_e/\text{cm}^{-3})}{(v_r/\alpha c)^9}.
\end{equation}

Finally, the total capture rate is obtained as,
\begin{equation}
C_q = \alpha_{REC} + \alpha_{DR} + \alpha_{3BR},
\label{ec_total_capture_rate}
\end{equation}
evaluated for the charge state $q$.

\begin{figure}[]
	\centering
		\includegraphics[width=15cm]{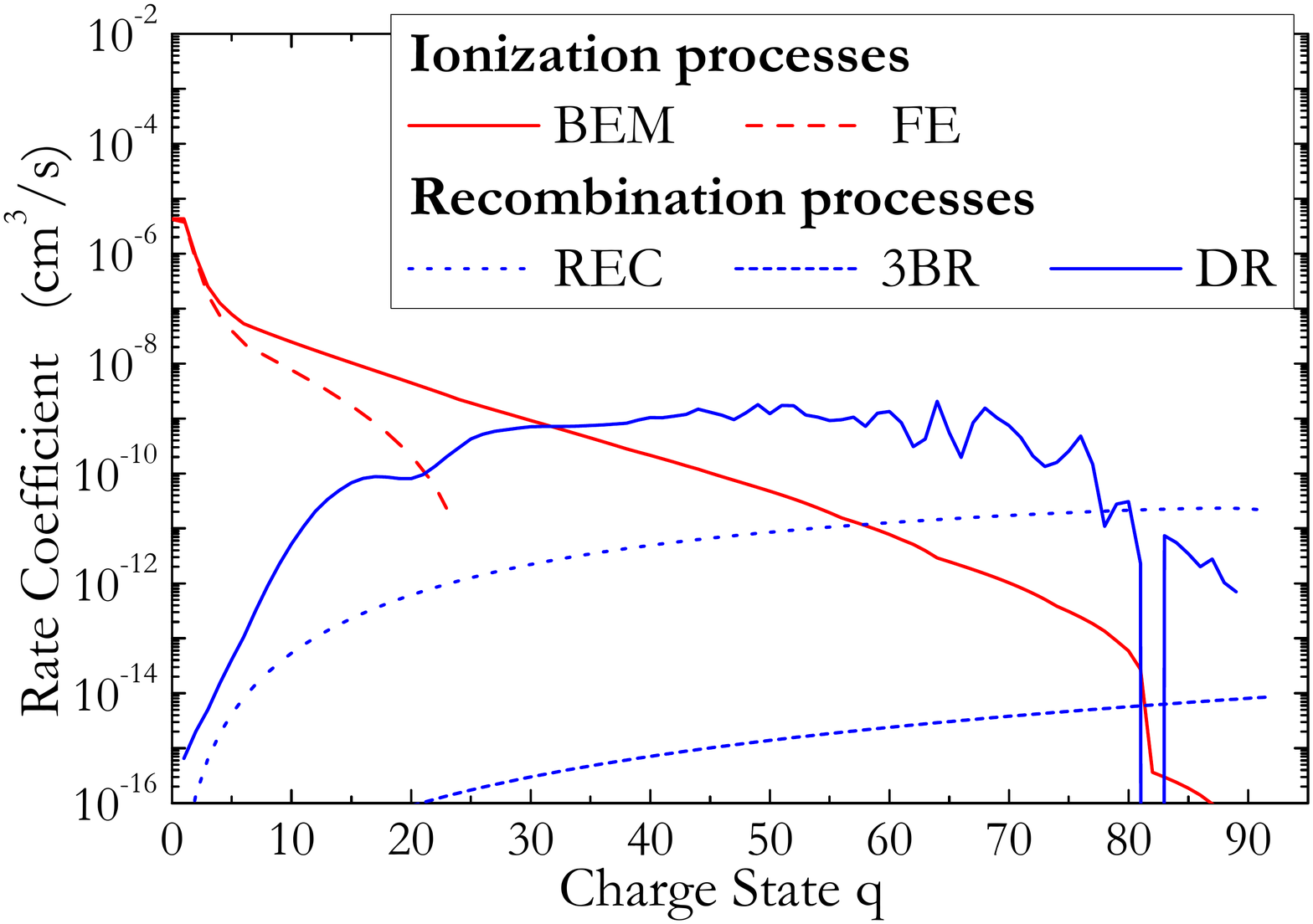}
		\caption{Ionization and capture rates for uranium ions ($E = 1.4$ MeV/u) in a hydrogen plasma ($T=2$ eV, $n_e\simeq 10^{17}$ cm$^{-3}$).}
		\label{fig_U_H_cross_sections_1}
\end{figure}	

Figure \ref{fig_U_H_cross_sections_1} shows the rates of all ionization and recombination processes for the case of an uranium beam at $1.4$ MeV/u energy traveling through a fully ionized hydrogen plasma at $T = 2$ eV in units of cm$^3/$s (density independent). As can be seen, the dielectronic recombination (DR) is the dominant process for the projectile electron capture. Also, the shell structure of the DR is clearly observed. Moreover, the projectile ionization in collisions with plasma ions is the dominant electron loss process. Also, the ionization edge, which means that the free electrons with $v_r < v_n$ do not have a sufficient kinetic energy for the ionization of a projectile bound electron in the $n_{th}$ shell,  is confirmed.
	

\subsection{Method II. Analytical  model}\label{sec_transitory_number_bound_electrons}

\subsubsection{Kreussler's criterion}
		
One of the most used procedures to determine the equilibrium charge state of heavy ions is based on the Bohr's criterion \cite{Bohr48}, which was subsequently modified by Kreussler \textit {et al.} \cite{KreusslerPRB81}, who proposed that the equilibrium charge state depends on the relative velocity between the projectile, with velocity $ v $, and the target electrons, with velocity $ v_e $. The relative velocity between the projectile and the target electrons averaged over all possible orientations of the vector $ \mathbf {v} - \mathbf {v_e} $ is given by
\begin{equation}
v_{rk}=\left| \mathbf{v} - \mathbf{v_e} \right| = \frac{v_e^2}{6v} \left[ \left( \frac{v}{v_e} + 1 \right)^3 - \left| \frac{v}{v_e} - 1 \right|^3 \right].
\label{ec_vr}
\end{equation}

If the target is a plasma, the target electron velocity depends on the Fermi velocity, $v_F$, as in solids, and also on the thermal velocity of the plasma free electrons, $ v_ {the}$. Considering both contributions, the velocity of the plasma electrons is given by
\begin{equation}
v_e  = \left(2 \frac{3}{5} E_F + 3 k_BT \right)^{1/2}
\label{ec_ve}
\end{equation}

Then, the equilibrium charge state given by the Kreussler's criterion can be obtained as,
\begin{equation}
Q_{eq} = Z - N_{eq} = Z  - Ze^{- v_{rk} / Z^{2/3} v_0},
\label{eq_Kreussler_criterion}
\end{equation}
where $N_{eq}$ is the equilibrium number of bound electrons and $ Z^{2/3} v_0 $ is the velocity of the projectile bound electrons in the Thomas-Fermi model, $v_0$ being the Bohr's velocity. The equilibrium charge state increases with the relative velocity between projectile and target electrons until it reaches its high-velocity limit value $ Q_ {eq} = Z $.

		
\subsubsection{Peter's criterion}\label{sec_Peter_equation}
In the 40's, Bohr \cite{Bohr48, Bohr40, Bohr41}, and independently, Lamb \cite{Lamb40}, suggested the following criterion: in a sufficiently thick target the ionization and recombination processes balance out each other if the projectile velocity, $ v $, is equal to the orbital velocity, $  v_n $, of the most loosely bound electron of the projectile, i.e.,
$$
q=Q_{eq} \Leftrightarrow x_v \equiv \frac{v}{v_n} = 1.
$$
Later, Peter modified the above criterion, known as Bohr-Lamb criterion, by the relation \cite{PeterLPB90}
\begin{equation}
q=Q_{eq} \Leftrightarrow L_s(x_v)=C_s(x_v)
\label{ec_Bohr-Lamb-modificada}
\end{equation}
where $ s $ is the charge state with equal ionization and recombination rates and now $x_v \neq 1$. Here $ C_s $ is given by equation (\ref{equation_alpha_REC}) and $L_s$ is given by 
\begin{equation}
L_s (x_v)= 3\, a_0^2 \alpha \, c\, n_t \frac{Z_t^2}{v^5}\, q^2 \frac{x_v^8}{1+x_v^4}.
\label{ec_ls-aproximada}
\end{equation}

Substituting the equations (\ref{equation_alpha_REC}) and (\ref{ec_ls-aproximada}) in equation (\ref{ec_Bohr-Lamb-modificada}), the parameter $x_v$ can be solved as,
\begin{equation}
x_v=\left[2.21\times 10^{-6}\left(v^4/Z_t\right)\right]^{1/7.7}.
\label{eq_x_Peter2}
\end{equation}

Using equations (\ref{eq_x_Peter}) and (\ref{eq_x_Peter2}), the equilibrium charge state of an ion beam traveling through a fully ionized plasma given by the Peter's criterion can be obtained as, 
\begin{eqnarray}
Q_{eq} & = & \frac{3Z}{2}\mu^{1/3} \nonumber \\
& & \times \left[\left(\sqrt{1+\mu}+1\right)^{\frac{1}{3}}-\left(\sqrt{1+\mu}-1\right)^{\frac{1}{3}}\right],
\label{eq_Peter_criterion}
\end{eqnarray}
where $\mu=35.5 v^{1.44}Z^{-2}Z_t^{0.39}$ and $v$ is in atomic units. Here, only the electron loss in collisions with the plasma ions and the electron capture by means of the radiative electron capture have been taken into account. 

\subsubsection{Analytical model for the instantaneous charge state}\label{sec_analytical_instantaneous_charge_state}
For an ion beam with initial charge state $Q_0$ traveling with velocity $v$ through a plasma, the instantaneous charge state can estimated as,
\begin{equation}
Q(x) = Q_{eq} - \left( Q_{eq} - Q_0 \right) \text{exp} \left (-\frac{x}{\lambda_{ion}} \right),
\label{eq_analytical_model}
\end{equation}
where $Q_{eq}$ is the equilibrium charge state according to the Kreussler's criterion, [Eq. (\ref{eq_Kreussler_criterion})], or the one obtained with the Peter's criterion [Eq. (\ref{eq_Peter_criterion})], $x$ is the depth (or plasma length) traveled by the projectile and $\lambda_ {ion} $ is the ionization length which can be estimated by 
\begin{equation}
\lambda_{ion}=\frac{1}{n_e \sigma_{ion}},
\label{ec_lambda_ion_1}
\end{equation}
where $\sigma_{ion}$  is the ionization cross section. In the case of cold matter, $\sigma_{ion} \simeq 10^{-17}$ cm$^2$ \cite{TakamotoNIMB99}, so that the necessary length (or time) to reach the equilibrium charge state is mainly a function of the projectile velocity. 

\begin{figure}[]
	\centering
		\includegraphics[width=15cm]{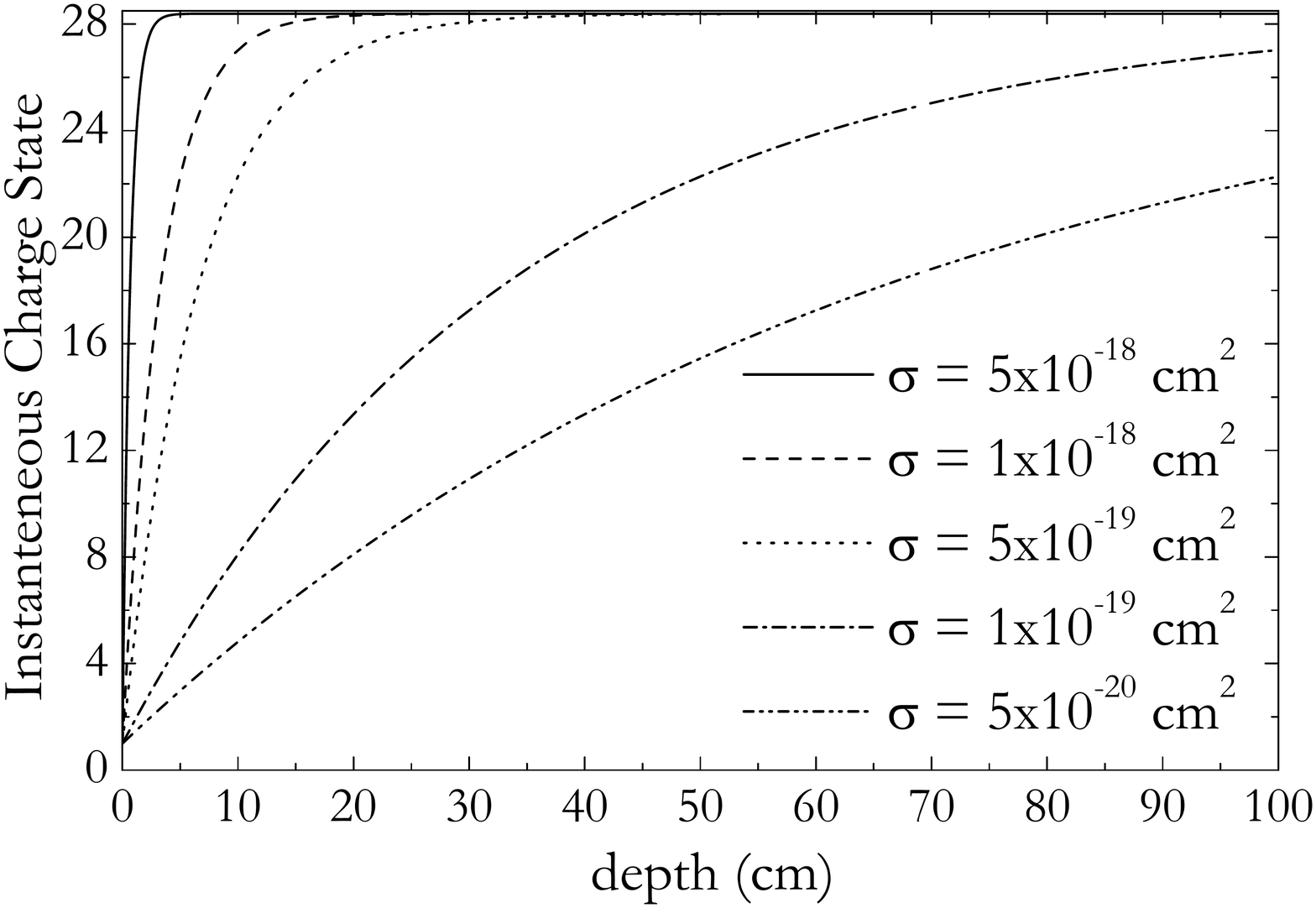}
		\caption{Instantaneous charge state [Eq. (\ref{eq_analytical_model})] of uranium ions ($E = 1.4$ MeV/u) in a hydrogen plasma ($T=2$ eV, $n_e = 3 \times 10^{17}$ cm$^{-3}$) for different values of $\sigma_{ion}$.}
		\label{fig_sigma_dependence}
\end{figure}

Figure \ref{fig_sigma_dependence} shows an illustrative example of the strong influence of the ionization cross section, $\sigma_{ion}$, on the instantaneous charge state given by equation (\ref{eq_analytical_model}). The ionization cross section is determined by the necessary time for the ion to reach its equilibrium charge state,
\begin{equation}
\sigma_{ion} = \frac{\nu_{eq}}{n_e v},
\label{ec_sigma_ion}
\end{equation} 
where $\nu_{eq} = 1 / t_{eq}$, $t_{eq}$ being the necessary time to achieve the equilibrium charge state. Substituting equation (\ref{ec_sigma_ion}) in equation (\ref{ec_lambda_ion_1}), the ionization length is given by
\begin{equation}
\lambda_{ion}=\frac{v}{\nu_{eq}}.
\label{ec_lambda_ion_2}
\end{equation}

\begin{figure}[]
	\centering
		\includegraphics[width=15cm]{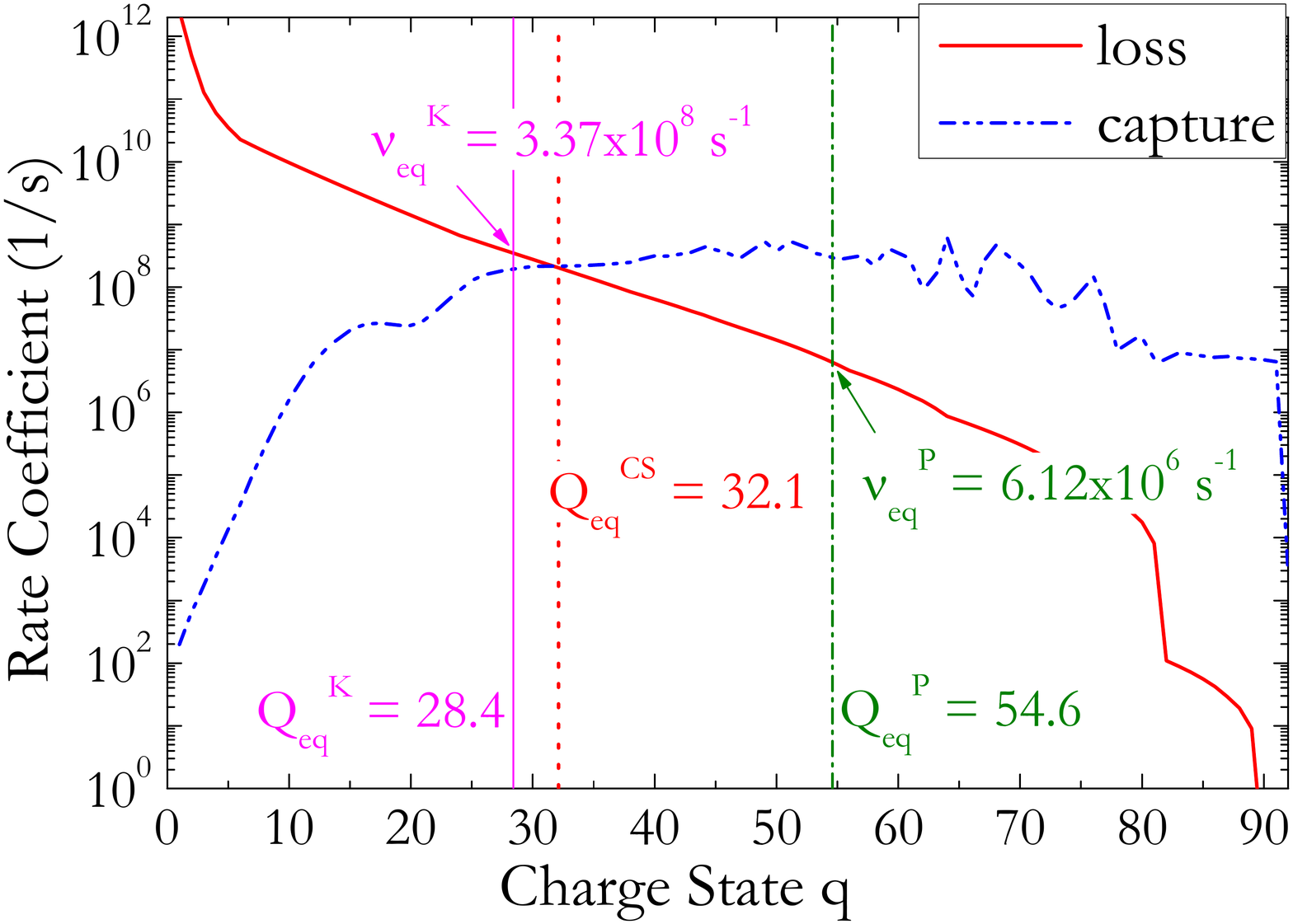}
		\caption{Ionization and capture rates for uranium ions ($E = 1.4$ MeV/u) in a fully ionized hydrogen plasma ($T=2$ eV, $n_e = 3 \times 10^{17}$ cm$^{-3}$). Full line: electron loss rate; Dashed dotted dotted line: electron capture rate; Vertical lines indicate the equilibrium charge state according to: Kreussler's criterion [Eq. (\ref{eq_Kreussler_criterion})] (full line), cross sections model [Eq. (\ref{eq_eq_cross_sections_model})] (dotted line) and Peter's criterion [Eq. (\ref{eq_Peter_criterion})] (dashed dotted line).}
		\label{fig_U_H_cross_sections_2}
\end{figure}

Figure \ref{fig_U_H_cross_sections_2} shows the importance of the parameter $\nu_{eq} $. Here, the ionization and recombination rates are estimated, in units s$^{-1}$ (density dependent), for the case of an uranium beam at $1.4$ MeV/u energy, traveling through a fully ionized hydrogen plasma at $T = 2$ eV and $n_e = 3 \times 10^{17}$ cm$^{-3}$. $\nu_{eq}$ is the ionization rate evaluated at the equilibrium charge state. Due to the different equilibrium charge state given by both the Kreussler's and the Peter's criteria, $\nu_{eq}$ is also different. 

In this work, $\nu_{eq}$ has been calculated from the sum of the ionization with plasma ions and the ionization with plasma free electrons, i.e., by means of the equation (\ref{ec_total_loss_rate}) evaluated at the equilibrium charge state according to the Kreussler's criterion, $Q_{eq}^K = 28.4$ and $\nu_{eq}^K = 3.37 \times 10^6$ s$^{-1}$, and at the equilibrium charge state obtained with the Peter's criterion, $Q_{eq}^K = 54.6$ and $\nu_{eq}^P = 6.12 \times 10^8$ s$^{-1}$. 

If the ionization rates are unknown, the equation (\ref{ec_ls-aproximada}) can be applied to estimate $\nu_{eq}$. However, this expression is an approximation and therefore, it should be used only for rough estimations.


\section{Results}\label{sec_results}

The validity of the charge state models described in section \ref{sec_Projectile_charge_state} is checked by comparing the theoretical predictions with experimental data for energy loss obtained from one of the first experiments carried out in the 90's at GSI-Darmstadt by Hoffmann \textit{et al.} \cite{HoffmannPRA90}. 

In the analyzed experiment, an uranium beam with initial charge state $Q_0 = 33$ is accelerated to an energy of $ 1.4$ MeV/u in the UNILAC accelerator. The plasma is designed as a linear discharge with z-pinch geometry using hydrogen gas. This gas is confined in a quartz tube of 4 cm in diameter and 36 cm in length. At both ends, the tube is closed with metal electrodes having apertures of 4 cm in diameter to allow the entrance and exit of the beam. The plasma conditions, described by the free electron density, $ n_e $, and the plasma temperature, $ T $, are measured by monitoring the light emission from the plasma, mainly the emission of the $ H_\beta $ line. The measurement of the energy loss of heavy ions is performed by the time-of-flight method (TOF).

	\begin{figure}[]
	\centering
		\includegraphics[width=15cm]{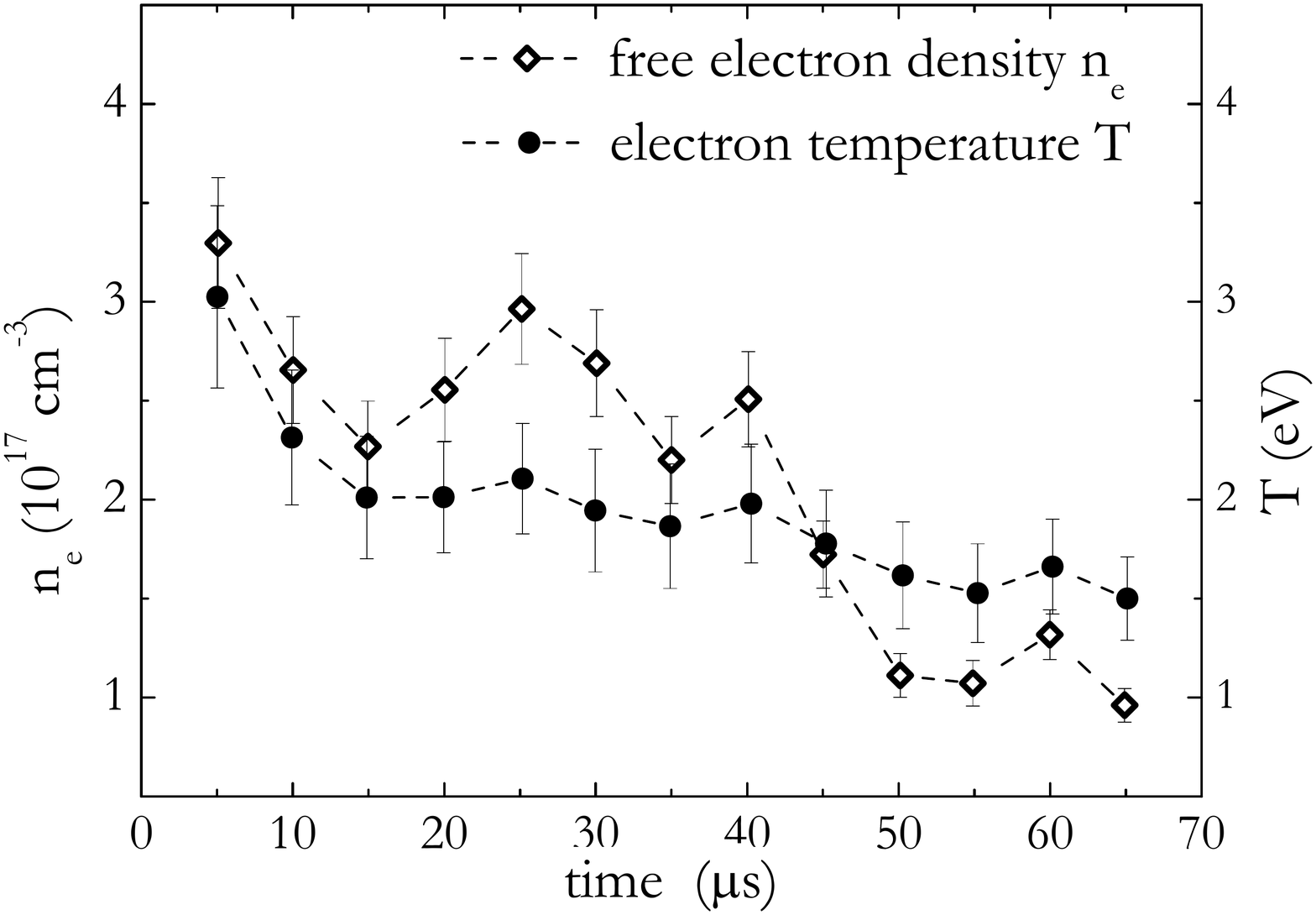}
		\caption{Variation of the measured plasma parameters, from Ref. \cite{HoffmannPRA90}, free-electron density, $n_e$, and temperature, $T$, with time.}
		\label{fig_plasma_conditions}
\end{figure}

Figure \ref{fig_plasma_conditions} shows the measured time evolution of the plasma conditions. The plasma conditions at \mbox{$t = 25$ $\mu$s}, i.e., $n_e=3\times 10^{17}$ cm$^{-3}$ and $T =2$ eV, have been chosen to conduct this work because the free electron density reaches its maximum, which means the maximum energy loss of uranium ions at this time. Here, it is important to note that the TOF of the uranium ions at $1.4$ MeV/u for a distance $l=36$ cm is much smaller than the time evolution of the plasma conditions.


	\subsection{Energy loss model}\label{sec_energy_loss_model}
The energy loss model, briefly described in the following, has been widely studied by our research group \cite{BarrigaPRE10, BarrigaPRE16}. In atomic units, the electronic stopping in the Random Phase Approximation (RPA) description is defined as,
\begin{eqnarray}
-\frac{dE}{dx}\equiv S_e =\frac{2}{\pi v^2} \int_0^{\infty} {dk \frac{\left[Z-\rho_e(k)\right]^2}{k}} \nonumber\\ 
\times \int_0^{kv} {dw w\, \text{Im} \left[\frac{-1}{\epsilon(k,w)}\right]}.
\label{eq_stopping_power}
\end{eqnarray}

The electronic stopping depends on the target conditions through the so-called Energy Loss Function (ELF), which is given by
$$
ELF=\text{Im} \left(\frac{1}{\epsilon (k,w)}\right),
$$
where $\epsilon(k,w)$ is the dielectric function of the target \cite{BarrigaPRE10,BarrigaPRE16}. 

The $\rho_e(k)$ is the Fourier transform of the electron density distribution of the projectile in the Brandt-Kitagawa (BK) model given by \cite{BrandtPRB82}
\begin{equation}
\rho_e(k)=\frac{N}{1+\left(k\Lambda\right)^2},
\label{ec_rho_e}
\end{equation}
where $N$ is the number of electrons bound to the projectile nucleus and $\Lambda$ is a variational parameter given by
\begin{equation}
\Lambda=\frac{0.48 N^{2/3}}{Z - \frac{1}{7} N}.
\end{equation}

An instantaneous electron density distribution can be estimated by including the depth dependence of the number of bound electrons,
\begin{equation}
\rho_e(k,x) = \frac{ N(x) }{1 + \left( k \Lambda(x) \right)^2},
\label{eq_rho_e_x}
\end{equation}
where $N(x) = Z - Q(x)$ is the instantaneous number of bound electrons.

Here, the strong dependence of the plasma stopping power with the projectile charge state [$\left(Z - \rho_e(k) \right)^2$ in equation (\ref{eq_stopping_power})] can be used to check the validity of the charge state models described in section \ref{sec_Projectile_charge_state} by comparing the theoretical predictions with energy loss measurements. 

Most authors evaluate the plasma stopping power assuming that the projectile equilibrium charge state during the whole plasma length. However, it must be considered that the charge state of the projectile could be different from its equilibrium value. Therefore, the theoretical energy loss should take into account the instantaneous charge state of the projectile. 

In this model, the energy loss as a function of the depth traveled by the projectile is given by
\begin{eqnarray}
S_e(x)= \frac{2}{\pi v^2} \int_0^{\infty} {dk \frac{\left[Z-\rho_e(k,x)\right]^2}{k}} \nonumber\\ 
\times \int_0^{kv} {dw w Im\left[\frac{-1}{\epsilon(k,w)}\right]}.
\label{eq_stopping_power_x}
\end{eqnarray}

Now, all instantaneous charge state models studied in this work can be taken into account, and the total electronic stopping along the propagation in the plasma is then obtained as,
\begin{equation}
S_e = \frac{1}{L_p} \int_0^{L_p} S_e(x) dx,
\label{eq_mean_stopping_power}
\end{equation}
where $L_p$ is the total plasma length and $S_e(x)$ is given by equation (\ref{eq_stopping_power_x}).

	\subsection{Charge state}\label{sec_charge_state_results}
	
\begin{figure}[]
	\centering
		\includegraphics[width=15cm]{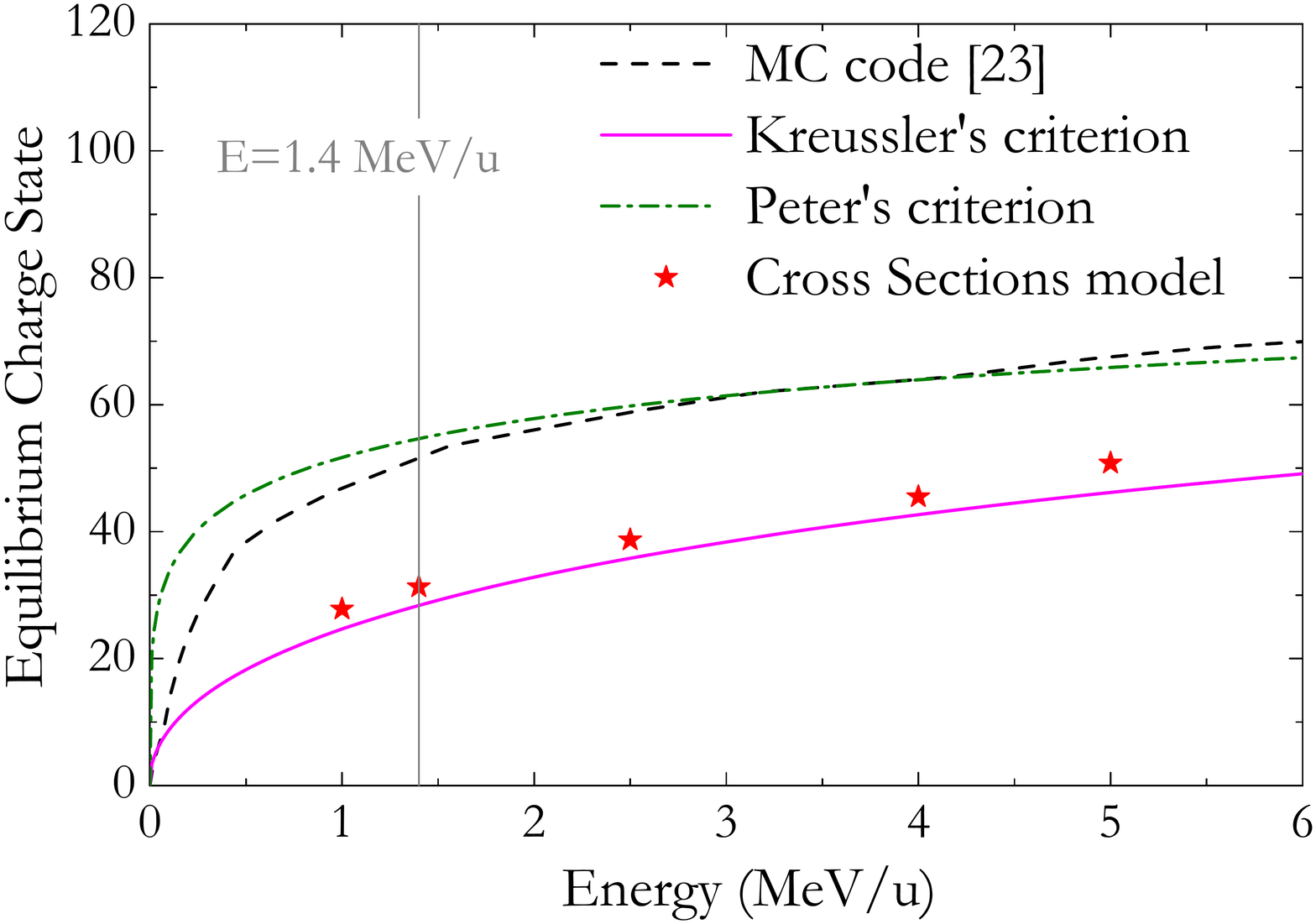}
		\caption{Equilibrium charge state of uranium ions in a fully ionized hydrogen plasma at $T=2$ eV and $n_e=3\times 10^{17}$ cm$^{-3}$. Dashed line: Monte Carlo code from  Ref. \cite{HoffmannPRA90}; full line: Kreussler's criterion [Eq. (\ref{eq_Kreussler_criterion})]; dashed dotted line: Peter's criterion [Eq. (\ref{eq_Peter_criterion})]; symbols: cross sections model [Eq. (\ref{eq_eq_cross_sections_model})]; the vertical line indicates the energy of the experiment.}
		\label{fig_equilibrium_charge_state}
\end{figure}

Figure \ref{fig_equilibrium_charge_state} shows the equilibrium charge state according to all models described in section \ref{sec_Projectile_charge_state}. The results of the Monte Carlo (MC) code described in Ref. \cite{HoffmannPRA90} are also shown for comparison. As can be seen, the Peter's criterion fits well to the MC code used by Hoffmann \textit{et al.} except at low velocities, which is not the validity of the velocity range of the Peter's criterion. The reason of the good agreement at high velocity is that in both, the Peter's criterion and their MC code, the radiative electron capture is assumed to be the dominant electron capture process. This assumption lead to a much higher projectile ionization in plasma targets than in solid ones.

On the other hand, the results obtained with the Kreussler's criterion fits better to the cross sections model presented in this work. However, the first one is always slightly lower. The reason is that if the dielectronic recombination is the dominant electron capture process, the equilibrium charge state is not as high as Peter's criterion and MC code predict, and the Kreussler's criterion agrees with this behavior.

\begin{figure}[]
	\centering
		\includegraphics[width=15cm]{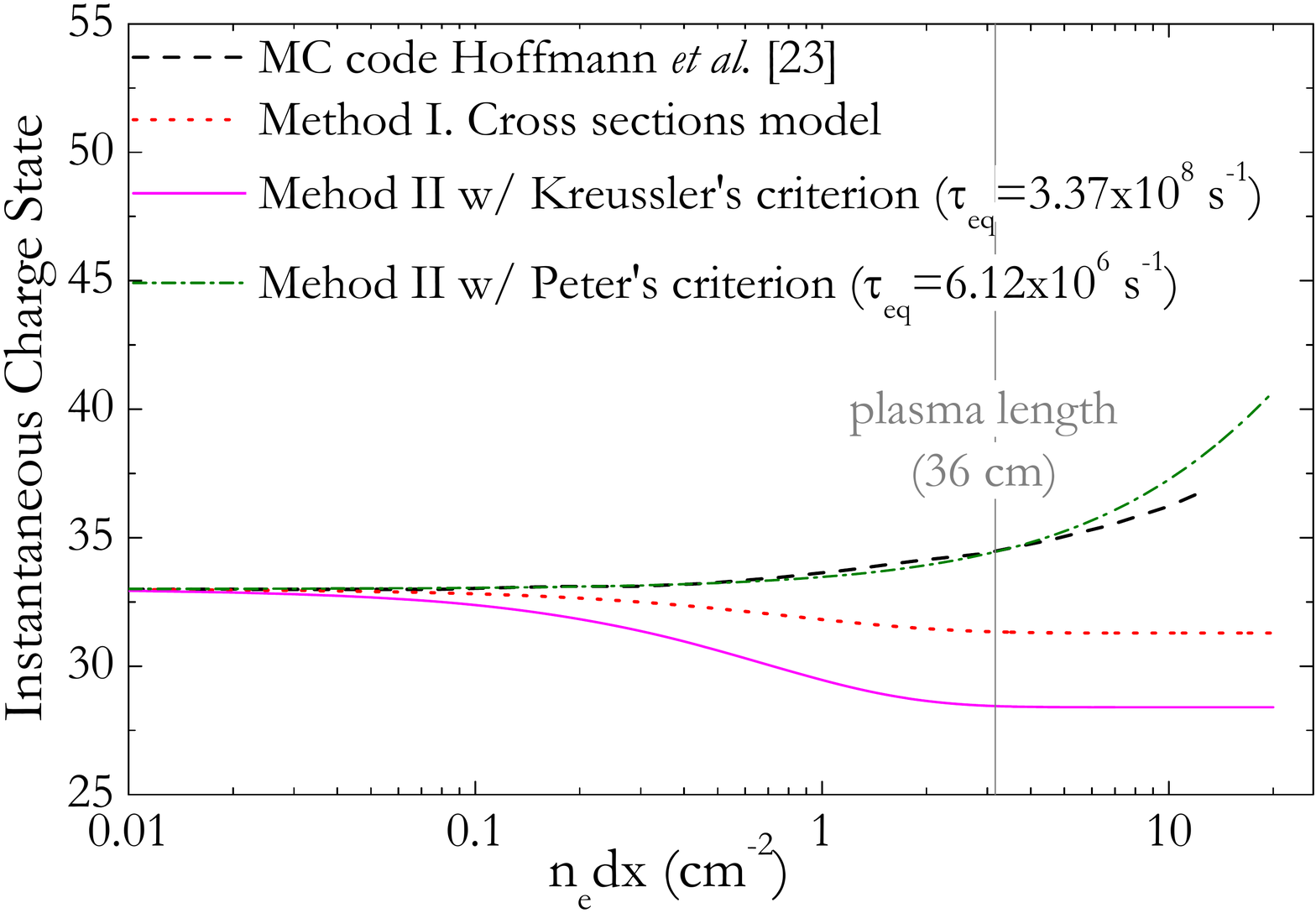}
		\caption{Instantaneous charge state of uranium ions ($E = 1.4$ MeV/u, $Q_0 = 33$) in a hydrogen plasma ($2$ eV, $10^{17}$ cm$^{-3}$). Dashed line:  MC code from Ref. \cite{HoffmannPRA90}; full line: Kreussler's criterion [Eqs. (\ref{eq_Kreussler_criterion}) and (\ref{eq_analytical_model})]; dashed dotted line: Peter's criterion [Eqs. (\ref{eq_Peter_criterion}) and (\ref{eq_analytical_model})]; dotted line: cross sections model [Eq. (\ref{eq_instantaneous_charge_state_cross_sections_model})]; the vertical line indicates the plasma length.}
		\label{fig_charge_state_evolution_U_H}
\end{figure}

In Figure \ref{fig_charge_state_evolution_U_H} the instantaneous charge state as a function of the areal density (or traveled depth) is showed according to all models described in section \ref{sec_Projectile_charge_state}, for an uranium ion beam, with initial charge state $Q_0 = 33$ at an energy of $1.4$ MeV/u, impinge on a fully ionized hydrogen plasma at $T = 2$ eV and $n_e = 10^{17} $ cm$^{-3}$. The MC code results from Ref. \cite{HoffmannPRA90} are also shown for comparison. 

Here, the Peter's criterion shows the same behaviour as the MC code employed by Hoffmann \textit{et al.} in their work. Both models predict an increase of the projectile ionization inside the plasma. However, in both cases the instantaneous charge states are very different from the equilibrium ones as the value according to the Peter's criterion shows ($Q_{eq}^{P} = 54.6$). Due to the high projectile velocity and low target density, the necessary time (or depth) to reach the equilibrium charge, given by $\nu_{eq}^P = 6.12 \times 10^{6}$ s$^{-1}$, is very long and the projectile does not reach it.

On the other hand, the Kreussler's criterion has a similar evolution as the cross sections model. However, the Kreussler's criterion predicts a smaller instantaneous charge state due to the smaller equilibrium value, $Q_{eq}^K = 28.4$, than in the case of the cross sections model, $Q_{eq}^{CS} = 32.1$. Here, the necessary time to achieve equilibrium is similar in both cases due to the very similar $\nu_{eq} \simeq 3.3 \times 10^8$ s$^{-1}$.

As can be seen, the instantaneous charge state according to the Peter's criterion is larger than according to our cross sections model, while the one obtained with the Kreussler's criterion is smaller. However, the models do not differ as much as Figure \ref{fig_equilibrium_charge_state} predicted, showing that using the equilibrium charge state in any calculation that depends on it could lead to inaccuracies. 

Due to the better agreement in the prediction of the equilibrium charge state of the Kreussler's criterion with the cross sections model than when using the Peter's criterion, the first one should be used if the equilibrium charge state is achieved inside the plasma.

In view of the different equilibrium charge states predicted by the different models, significant differences in the projectile energy loss are expected, as its energy loss depends quadratically on its charge state [Eq. (\ref{eq_stopping_power})], as the experiment is in the linear interaction regime.

	\subsection{Comparison of theory with experimental data}\label{sec_energy_loss_results}
	
\begin{figure}[]
	\centering
		\includegraphics[width=15cm]{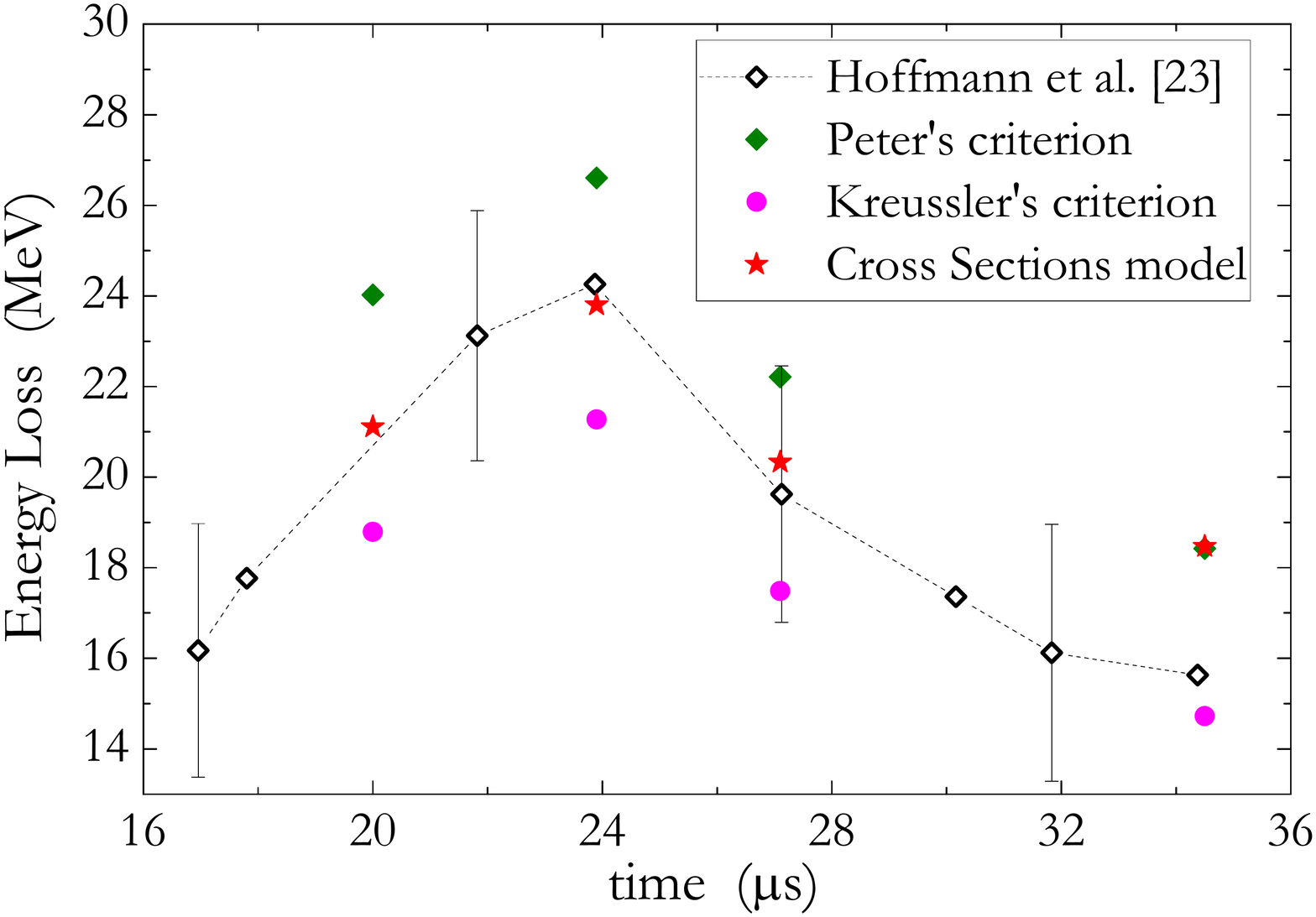}
		\caption{Energy loss of uranium ions ($E = 1.4$ MeV/u, $Q_0 = 33$) in a hydrogen plasma ($T\approx 2$ eV, $n_e\approx 10^{17}$ cm$^{-3}$). The charge state has been estimated according to all models described in section \ref{sec_Projectile_charge_state}. Opened diamonds: experiment from Ref. \cite{HoffmannPRA90}; closed diamonds: Peter's criterion; closed circles: Kreussler's criterion; closed stars: cross sections model.}
		\label{fig_energy_loss_1}
\end{figure}

In figure \ref{fig_energy_loss_1}, our theoretical predictions are compared with experimental data ​​of an uranium beam, with initial charge state $Q_0 = 33$ and energy $ 1.4 $ MeV/u, traveling through a fully ionized hydrogen plasma. The time evolution of plasma parameters is shown in Figure \ref{fig_plasma_conditions} ($T\approx 2$ eV and $n_e\approx10^{17}$ cm$^{-3}$). The energy loss has been calculated considering the instantaneous charge state according to all charge state models described in section \ref{sec_rate_equations} and showed in Figure \ref{fig_charge_state_evolution_U_H}. As can be seen, predictions using the cross sections model fit very well to experimental data, which suggests that the instantaneous charge state predicted by this model is accurate. The disagreement between our theoretical prediction with the cross sections model and experimental data at $t \simeq 34$ $\mu$s could be due to that at this time the plasma is not fully ionized as our ideal model supposes, which implies that the projectile charge state could be smaller due to the capture of target bound electron.

On the other hand, predictions using the analytical equilibrium model overestimate, in the case of using the charge state obtained with the Peter's criterion, or underestimate, in the case of using the one obtained with the Kreussler's criterion, the experimental energy loss. This is expected in view of the results shown in Figure \ref{fig_charge_state_evolution_U_H}. However, the analytical model can be applied for the energy loss prediction with a 15-20\% approximation.

Results show that both the Peter's criterion and the Hoffmann \textit{et al.} MC code do not fit well to reality due to the increasing of charge state inside the plasma predicted, while our cross sections model predicts a slightly reduction of the charge state with very good agreement with experimental data. However, experimental data in conditions where the equilibrium charge state inside the plasma is achieved (lower velocities and/or higher densities and/or longer plasma) are necessary to confirm this prediction.

\begin{figure}[]
	\centering
		\includegraphics[width=15cm]{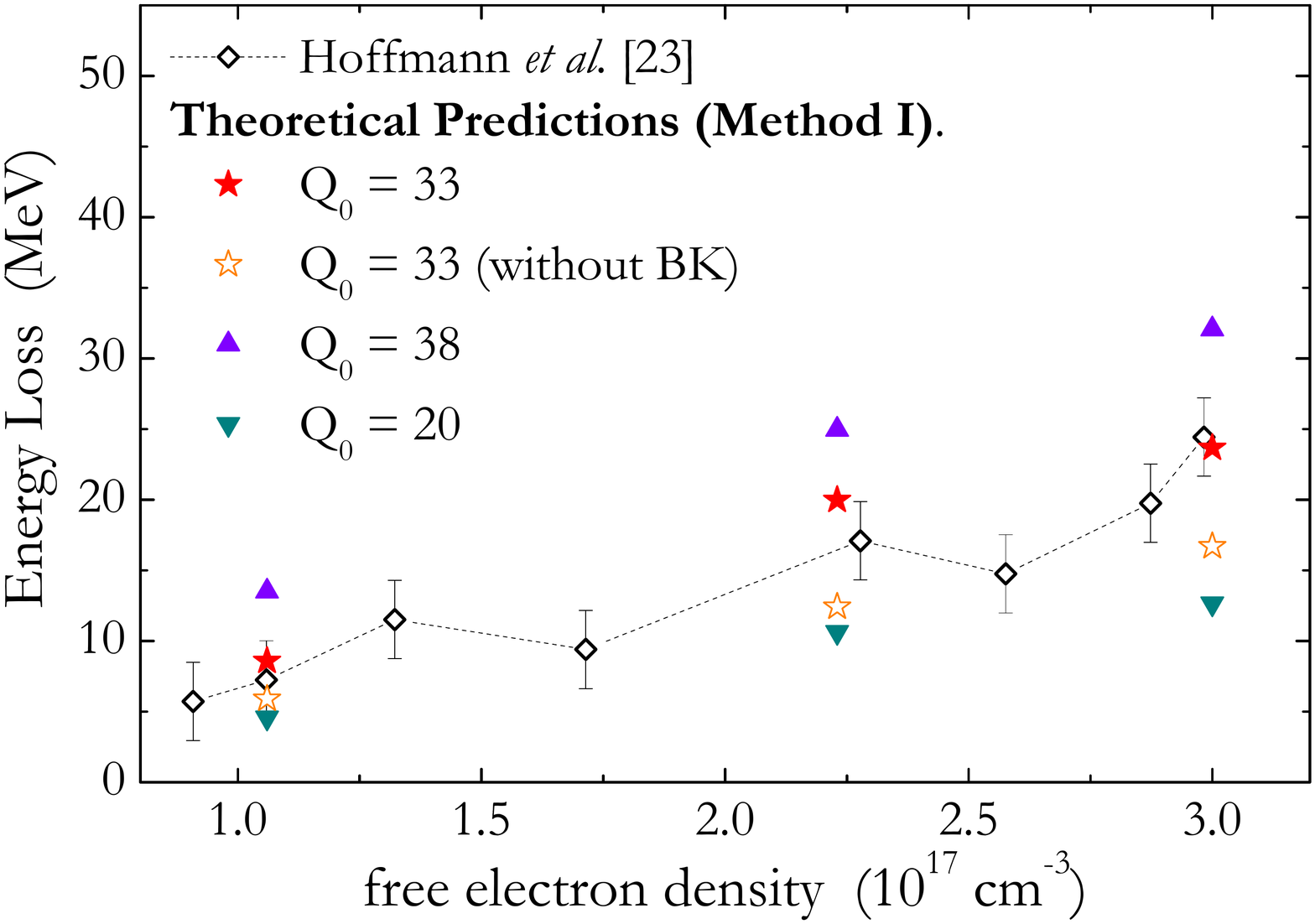}
		\caption{Energy loss of uranium ions ($E = 1.4$ MeV/u) in a hydrogen plasma. The charge state has been estimated according to the cross sections model [Eq. (\ref{eq_instantaneous_charge_state_cross_sections_model})]. Closed diamonds: experiment from Ref. \cite{HoffmannPRA90}; closed stars: with BK and $Q_0 = 33$; opened stars: without BK and $Q_0 = 33$; closed triangles up: with BK and $Q_0 = 38$; closed triangles down: with BK and $Q_0 = 20$.}
		\label{fig_energy_loss_2}
\end{figure}

Figure \ref{fig_energy_loss_2} shows the importance of including the Brandt-Kitagawa model [Eq. (\ref{eq_rho_e_x})] to describe the electron density distribution of the projectile in equation (\ref{eq_stopping_power}). The reason is that considering the electron density distribution of the projectile leads to a charge volume which increases the energy loss more than considering a point-like charge. Also, if the initial charge state used in the energy loss estimation is different to the one used for the authors, the theoretical estimations do not fit well to experimental data, showing that the initial charge state has an large influence in the energy loss. 

\section{Conclusions}\label{sec_conclusions}

In this work, two models to calculate the instantaneous charge state of heavy ions in fully ionized plasmas have been developed. In the first one, a detailed model including electron loss and capture cross sections has been used to determine the instantaneous charge state. In the second one, a simple analytical model employing equilibrium charge state formulas have been proposed to estimate the instantaneous charge state of the projectile. 

The equilibrium charge state has been also estimated using these electron loss and capture cross sections and compared with former equilibrium formulas. Here, the Kreussler's criterion agrees better with the cross sections model than the Peter's one.

Most authors use an equilibrium charge state instead of the instantaneous charge state in order to, for example, calculate the energy loss. However, whether the equilibrium charge state in plasmas is reached or not depends on the projectile velocity and on the target density, and assuming equilibrium along the whole propagating in the plasma can lead to inaccuracies in the energy loss estimation. For this reason, the instantaneous charge state as well as the initial charge state of the projectile must be taken into account in order to reproduce the experimental data.
 
It should be noted that the cross sections model is computationally time-consuming, while the analytical model is faster and only ionization cross sections, evaluated at the equilibrium charge state, are necessary. Here, the Kreussler's criterion rather than the Peter's criterion should be used if the dielectronic recombination is the dominant electron capture process and the equilibrium charge state is reached inside the plasma.

Finally, results show the necessity of using the Brandt-Kitagawa (BK) electronic density distribution in combination with the instantaneous charge state (taking into account the initial charge state) for a correct energy loss estimation. Comparing theory and experimental data, a very good agreement has been found with the model described in this work.

\section*{Acknowledgments}
R. M. would like to thank O. S. Haas and W. Cayzac for very helpful discussions. This work was financed by the LOEWE program of the Helmholtz International Center for FAIR [HIC4FAIR] (through a grant to R.M.). 


\bibliography{bibliography}

\begin{thebibliography}{59}%
\makeatletter
\providecommand \@ifxundefined [1]{%
 \@ifx{#1\undefined}
}%
\providecommand \@ifnum [1]{%
 \ifnum #1\expandafter \@firstoftwo
 \else \expandafter \@secondoftwo
 \fi
}%
\providecommand \@ifx [1]{%
 \ifx #1\expandafter \@firstoftwo
 \else \expandafter \@secondoftwo
 \fi
}%
\providecommand \natexlab [1]{#1}%
\providecommand \enquote  [1]{``#1''}%
\providecommand \bibnamefont  [1]{#1}%
\providecommand \bibfnamefont [1]{#1}%
\providecommand \citenamefont [1]{#1}%
\providecommand \href@noop [0]{\@secondoftwo}%
\providecommand \href [0]{\begingroup \@sanitize@url \@href}%
\providecommand \@href[1]{\@@startlink{#1}\@@href}%
\providecommand \@@href[1]{\endgroup#1\@@endlink}%
\providecommand \@sanitize@url [0]{\catcode `\\12\catcode `\$12\catcode
  `\&12\catcode `\#12\catcode `\^12\catcode `\_12\catcode `\%12\relax}%
\providecommand \@@startlink[1]{}%
\providecommand \@@endlink[0]{}%
\providecommand \url  [0]{\begingroup\@sanitize@url \@url }%
\providecommand \@url [1]{\endgroup\@href {#1}{\urlprefix }}%
\providecommand \urlprefix  [0]{URL }%
\providecommand \Eprint [0]{\href }%
\providecommand \doibase [0]{http://dx.doi.org/}%
\providecommand \selectlanguage [0]{\@gobble}%
\providecommand \bibinfo  [0]{\@secondoftwo}%
\providecommand \bibfield  [0]{\@secondoftwo}%
\providecommand \translation [1]{[#1]}%
\providecommand \BibitemOpen [0]{}%
\providecommand \bibitemStop [0]{}%
\providecommand \bibitemNoStop [0]{.\EOS\space}%
\providecommand \EOS [0]{\spacefactor3000\relax}%
\providecommand \BibitemShut  [1]{\csname bibitem#1\endcsname}%
\let\auto@bib@innerbib\@empty
\bibitem [{\citenamefont {Bell}(1953)}]{BellPR53}%
  \BibitemOpen
  \bibfield  {author} {\bibinfo {author} {\bibfnamefont {G.~I.}\ \bibnamefont
  {Bell}},\ }\href@noop {} {\bibfield  {journal} {\bibinfo  {journal} {Phys.
  Rev.}\ }\textbf {\bibinfo {volume} {90}},\ \bibinfo {pages} {548} (\bibinfo
  {year} {1953})}\BibitemShut {NoStop}%
\bibitem [{\citenamefont {Betz}(1983)}]{Betz83}%
  \BibitemOpen
  \bibfield  {author} {\bibinfo {author} {\bibfnamefont {H.~D.}\ \bibnamefont
  {Betz}},\ }\href@noop {} {\emph {\bibinfo {title} {Heavy Ion Charge
  States}}},\ Vol.~\bibinfo {volume} {4}\ (\bibinfo  {publisher} {Appl. At.
  Coll. Phys.},\ \bibinfo {year} {1983})\ p.~\bibinfo {pages} {1}\BibitemShut
  {NoStop}%
\bibitem [{\citenamefont {Bohr}(1948)}]{Bohr48}%
  \BibitemOpen
  \bibfield  {author} {\bibinfo {author} {\bibfnamefont {A.}~\bibnamefont
  {Bohr}},\ }\href@noop {} {\bibfield  {journal} {\bibinfo  {journal} {Mat.
  Fys. Medd. K. Dan.Vidensk. Selskab}\ }\textbf {\bibinfo {volume} {24}},\
  \bibinfo {pages} {1} (\bibinfo {year} {1948})}\BibitemShut {NoStop}%
\bibitem [{\citenamefont {Northcliffe}(1960)}]{NorthcliffePR60}%
  \BibitemOpen
  \bibfield  {author} {\bibinfo {author} {\bibfnamefont {L.~C.}\ \bibnamefont
  {Northcliffe}},\ }\href@noop {} {\bibfield  {journal} {\bibinfo  {journal}
  {Phys. Rev.}\ }\textbf {\bibinfo {volume} {120}},\ \bibinfo {pages} {1744}
  (\bibinfo {year} {1960})}\BibitemShut {NoStop}%
\bibitem [{\citenamefont {Kreussler}\ \emph {et~al.}(1981)\citenamefont
  {Kreussler}, \citenamefont {Varelas},\ and\ \citenamefont
  {Brandt}}]{KreusslerPRB81}%
  \BibitemOpen
  \bibfield  {author} {\bibinfo {author} {\bibfnamefont {S.}~\bibnamefont
  {Kreussler}}, \bibinfo {author} {\bibfnamefont {C.}~\bibnamefont {Varelas}},
  \ and\ \bibinfo {author} {\bibfnamefont {W.}~\bibnamefont {Brandt}},\
  }\href@noop {} {\bibfield  {journal} {\bibinfo  {journal} {Phys. Rev. B}\
  }\textbf {\bibinfo {volume} {23}},\ \bibinfo {pages} {82} (\bibinfo {year}
  {1981})}\BibitemShut {NoStop}%
\bibitem [{\citenamefont {Ziegler}\ \emph {et~al.}(1985)\citenamefont
  {Ziegler}, \citenamefont {Biersack},\ and\ \citenamefont {Littmark}}]{SRIM}%
  \BibitemOpen
  \bibfield  {author} {\bibinfo {author} {\bibfnamefont {J.~F.}\ \bibnamefont
  {Ziegler}}, \bibinfo {author} {\bibfnamefont {J.~P.}\ \bibnamefont
  {Biersack}}, \ and\ \bibinfo {author} {\bibfnamefont {U.}~\bibnamefont
  {Littmark}},\ }\href@noop {} {\emph {\bibinfo {title} {The Stopping and Range
  of Ions in Matter}}}\ (\bibinfo  {publisher} {Pergamon, New York},\ \bibinfo
  {year} {1985})\BibitemShut {NoStop}%
\bibitem [{\citenamefont {Bohr}\ and\ \citenamefont {Lindhard}(1954)}]{Bohr54}%
  \BibitemOpen
  \bibfield  {author} {\bibinfo {author} {\bibfnamefont {N.}~\bibnamefont
  {Bohr}}\ and\ \bibinfo {author} {\bibfnamefont {J.}~\bibnamefont
  {Lindhard}},\ }\href@noop {} {\bibfield  {journal} {\bibinfo  {journal} {Mat.
  Fys. Medd. K. Dan. Vidensk. Selskab}\ }\textbf {\bibinfo {volume} {28}}
  (\bibinfo {year} {1954})}\BibitemShut {NoStop}%
\bibitem [{\citenamefont {Betz}(1972)}]{Betz72}%
  \BibitemOpen
  \bibfield  {author} {\bibinfo {author} {\bibfnamefont {H.~D.}\ \bibnamefont
  {Betz}},\ }\href@noop {} {\bibfield  {journal} {\bibinfo  {journal} {Rev.
  Mod. Phys.}\ }\textbf {\bibinfo {volume} {44}},\ \bibinfo {pages} {465}
  (\bibinfo {year} {1972})}\BibitemShut {NoStop}%
\bibitem [{\citenamefont {Wittkower}\ and\ \citenamefont
  {Betz}(1973)}]{Wittkower-Betz1973}%
  \BibitemOpen
  \bibfield  {author} {\bibinfo {author} {\bibfnamefont {A.}~\bibnamefont
  {Wittkower}}\ and\ \bibinfo {author} {\bibfnamefont {H.-D.}\ \bibnamefont
  {Betz}},\ }\href@noop {} {\bibfield  {journal} {\bibinfo  {journal} {Phys.
  Rev. A}\ }\textbf {\bibinfo {volume} {7}},\ \bibinfo {pages} {159} (\bibinfo
  {year} {1973})}\BibitemShut {NoStop}%
\bibitem [{\citenamefont {Shima}\ \emph {et~al.}(1989)\citenamefont {Shima},
  \citenamefont {Kuno}, \citenamefont {Kakita},\ and\ \citenamefont
  {Yamanoucht}}]{Shima89}%
  \BibitemOpen
  \bibfield  {author} {\bibinfo {author} {\bibfnamefont {K.}~\bibnamefont
  {Shima}}, \bibinfo {author} {\bibfnamefont {N.}~\bibnamefont {Kuno}},
  \bibinfo {author} {\bibfnamefont {T.}~\bibnamefont {Kakita}}, \ and\ \bibinfo
  {author} {\bibfnamefont {M.}~\bibnamefont {Yamanoucht}},\ }\href@noop {}
  {\bibfield  {journal} {\bibinfo  {journal} {Phys. Rev. A}\ }\textbf {\bibinfo
  {volume} {39}},\ \bibinfo {pages} {4316} (\bibinfo {year}
  {1989})}\BibitemShut {NoStop}%
\bibitem [{\citenamefont {Rozet}\ \emph {et~al.}(1996)\citenamefont {Rozet},
  \citenamefont {St\'ephan},\ and\ \citenamefont {Vernhet}}]{RozetNIMB96}%
  \BibitemOpen
  \bibfield  {author} {\bibinfo {author} {\bibfnamefont {J.~P.}\ \bibnamefont
  {Rozet}}, \bibinfo {author} {\bibfnamefont {C.}~\bibnamefont {St\'ephan}}, \
  and\ \bibinfo {author} {\bibfnamefont {D.}~\bibnamefont {Vernhet}},\
  }\href@noop {} {\bibfield  {journal} {\bibinfo  {journal} {Nucl. Instrum.
  Meth. B}\ }\textbf {\bibinfo {volume} {107}},\ \bibinfo {pages} {67}
  (\bibinfo {year} {1996})}\BibitemShut {NoStop}%
\bibitem [{\citenamefont {Lamour}\ \emph {et~al.}(2015)\citenamefont {Lamour},
  \citenamefont {Fainstein}, \citenamefont {Galassi}, \citenamefont {Prigent},
  \citenamefont {Ramirez}, \citenamefont {Rivarola}, \citenamefont {Rozet},
  \citenamefont {Trassilleni},\ and\ \citenamefont {Vernhet}}]{LamourPRA15}%
  \BibitemOpen
  \bibfield  {author} {\bibinfo {author} {\bibfnamefont {E.}~\bibnamefont
  {Lamour}}, \bibinfo {author} {\bibfnamefont {P.~D.}\ \bibnamefont
  {Fainstein}}, \bibinfo {author} {\bibfnamefont {M.}~\bibnamefont {Galassi}},
  \bibinfo {author} {\bibfnamefont {C.}~\bibnamefont {Prigent}}, \bibinfo
  {author} {\bibfnamefont {C.~A.}\ \bibnamefont {Ramirez}}, \bibinfo {author}
  {\bibfnamefont {R.~D.}\ \bibnamefont {Rivarola}}, \bibinfo {author}
  {\bibfnamefont {J.~P.}\ \bibnamefont {Rozet}}, \bibinfo {author}
  {\bibfnamefont {M.}~\bibnamefont {Trassilleni}}, \ and\ \bibinfo {author}
  {\bibfnamefont {D.}~\bibnamefont {Vernhet}},\ }\href@noop {} {\bibfield
  {journal} {\bibinfo  {journal} {Phys. Rev. A}\ }\textbf {\bibinfo {volume}
  {92}},\ \bibinfo {pages} {042703} (\bibinfo {year} {2015})}\BibitemShut
  {NoStop}%
\bibitem [{\citenamefont {Shevelko}\ \emph {et~al.}(2010)\citenamefont
  {Shevelko}, \citenamefont {Stohlker}, \citenamefont {Tawara}, \citenamefont
  {Tolstikhina},\ and\ \citenamefont {Weber}}]{ShevelkoNIMB10}%
  \BibitemOpen
  \bibfield  {author} {\bibinfo {author} {\bibfnamefont {V.~P.}\ \bibnamefont
  {Shevelko}}, \bibinfo {author} {\bibfnamefont {T.}~\bibnamefont {Stohlker}},
  \bibinfo {author} {\bibfnamefont {H.}~\bibnamefont {Tawara}}, \bibinfo
  {author} {\bibfnamefont {I.~Y.}\ \bibnamefont {Tolstikhina}}, \ and\ \bibinfo
  {author} {\bibfnamefont {G.}~\bibnamefont {Weber}},\ }\href@noop {}
  {\bibfield  {journal} {\bibinfo  {journal} {Nucl. Instrum. Meth. B}\ }\textbf
  {\bibinfo {volume} {268}},\ \bibinfo {pages} {2611} (\bibinfo {year}
  {2010})}\BibitemShut {NoStop}%
\bibitem [{\citenamefont {Litsarev}(2013)}]{LitsarevCPC13}%
  \BibitemOpen
  \bibfield  {author} {\bibinfo {author} {\bibfnamefont {M.~S.}\ \bibnamefont
  {Litsarev}},\ }\href@noop {} {\bibfield  {journal} {\bibinfo  {journal}
  {Comput. Phys. Commun.}\ }\textbf {\bibinfo {volume} {184}},\ \bibinfo
  {pages} {432} (\bibinfo {year} {2013})}\BibitemShut {NoStop}%
\bibitem [{\citenamefont {Ogawa}\ \emph {et~al.}(1999)\citenamefont {Ogawa},
  \citenamefont {Neuner}, \citenamefont {Sakumi}, \citenamefont {Hasegawa},
  \citenamefont {Sasa}, \citenamefont {Horioka}, \citenamefont {Ogure},
  \citenamefont {Hattori}, \citenamefont {Shiho},\ and\ \citenamefont
  {Miyamoto}}]{OgawaFED99}%
  \BibitemOpen
  \bibfield  {author} {\bibinfo {author} {\bibfnamefont {M.}~\bibnamefont
  {Ogawa}}, \bibinfo {author} {\bibfnamefont {U.}~\bibnamefont {Neuner}},
  \bibinfo {author} {\bibfnamefont {A.}~\bibnamefont {Sakumi}}, \bibinfo
  {author} {\bibfnamefont {J.}~\bibnamefont {Hasegawa}}, \bibinfo {author}
  {\bibfnamefont {K.}~\bibnamefont {Sasa}}, \bibinfo {author} {\bibfnamefont
  {K.}~\bibnamefont {Horioka}}, \bibinfo {author} {\bibfnamefont
  {Y.}~\bibnamefont {Ogure}}, \bibinfo {author} {\bibfnamefont
  {T.}~\bibnamefont {Hattori}}, \bibinfo {author} {\bibfnamefont {M.~S.}\
  \bibnamefont {Shiho}}, \ and\ \bibinfo {author} {\bibfnamefont
  {S.}~\bibnamefont {Miyamoto}},\ }\href@noop {} {\bibfield  {journal}
  {\bibinfo  {journal} {Fusion Eng. Des.}\ }\textbf {\bibinfo {volume} {44}},\
  \bibinfo {pages} {279} (\bibinfo {year} {1999})}\BibitemShut {NoStop}%
\bibitem [{\citenamefont {Eliezera1a}\ \emph {et~al.}(2007)\citenamefont
  {Eliezera1a}, \citenamefont {Murakamia},\ and\ \citenamefont
  {Vala}}]{EliezerLPB07}%
  \BibitemOpen
  \bibfield  {author} {\bibinfo {author} {\bibfnamefont {S.}~\bibnamefont
  {Eliezera1a}}, \bibinfo {author} {\bibfnamefont {M.}~\bibnamefont
  {Murakamia}}, \ and\ \bibinfo {author} {\bibfnamefont {J.~M.}\ \bibnamefont
  {Vala}},\ }\href@noop {} {\bibfield  {journal} {\bibinfo  {journal} {Laser
  Part. Beams}\ }\textbf {\bibinfo {volume} {25}},\ \bibinfo {pages} {585}
  (\bibinfo {year} {2007})}\BibitemShut {NoStop}%
\bibitem [{\citenamefont {Hoffmann}\ \emph {et~al.}(2007)\citenamefont
  {Hoffmann}, \citenamefont {Blazevic}, \citenamefont {Korosity}, \citenamefont
  {Ni}, \citenamefont {Pikuz}, \citenamefont {Rethfeld}, \citenamefont
  {Rosmej}, \citenamefont {Roth}, \citenamefont {Tahir}, \citenamefont {Udrea},
  \citenamefont {Varentsov}, \citenamefont {Weyrich}, \citenamefont {Sharkov},\
  and\ \citenamefont {Maron}}]{HoffmannNIMA07}%
  \BibitemOpen
  \bibfield  {author} {\bibinfo {author} {\bibfnamefont {D.~H.~H.}\
  \bibnamefont {Hoffmann}}, \bibinfo {author} {\bibfnamefont {A.}~\bibnamefont
  {Blazevic}}, \bibinfo {author} {\bibfnamefont {S.}~\bibnamefont {Korosity}},
  \bibinfo {author} {\bibfnamefont {P.}~\bibnamefont {Ni}}, \bibinfo {author}
  {\bibfnamefont {S.~A.}\ \bibnamefont {Pikuz}}, \bibinfo {author}
  {\bibfnamefont {B.}~\bibnamefont {Rethfeld}}, \bibinfo {author}
  {\bibfnamefont {O.}~\bibnamefont {Rosmej}}, \bibinfo {author} {\bibfnamefont
  {M.}~\bibnamefont {Roth}}, \bibinfo {author} {\bibfnamefont {N.~A.}\
  \bibnamefont {Tahir}}, \bibinfo {author} {\bibfnamefont {S.}~\bibnamefont
  {Udrea}}, \bibinfo {author} {\bibfnamefont {D.}~\bibnamefont {Varentsov}},
  \bibinfo {author} {\bibfnamefont {K.}~\bibnamefont {Weyrich}}, \bibinfo
  {author} {\bibfnamefont {B.~Y.}\ \bibnamefont {Sharkov}}, \ and\ \bibinfo
  {author} {\bibfnamefont {Y.}~\bibnamefont {Maron}},\ }\href@noop {}
  {\bibfield  {journal} {\bibinfo  {journal} {Nucl. Instrum. Meth. A}\ }\textbf
  {\bibinfo {volume} {577}},\ \bibinfo {pages} {8} (\bibinfo {year}
  {2007})}\BibitemShut {NoStop}%
\bibitem [{\citenamefont {Cook}\ \emph {et~al.}(2008)\citenamefont {Cook},
  \citenamefont {Kozioziemski}, \citenamefont {Nikroo}, \citenamefont
  {Wilkens}, \citenamefont {Bhandarkar}, \citenamefont {Forsman}, \citenamefont
  {Haan}, \citenamefont {Hoppe}, \citenamefont {Huang}, \citenamefont
  {Mapoles}, \citenamefont {Moody}, \citenamefont {Sater}, \citenamefont
  {Seugling}, \citenamefont {Stephens}, \citenamefont {Takagi},\ and\
  \citenamefont {Xu}}]{CookLPB08}%
  \BibitemOpen
  \bibfield  {author} {\bibinfo {author} {\bibfnamefont {R.}~\bibnamefont
  {Cook}}, \bibinfo {author} {\bibfnamefont {B.}~\bibnamefont {Kozioziemski}},
  \bibinfo {author} {\bibfnamefont {A.}~\bibnamefont {Nikroo}}, \bibinfo
  {author} {\bibfnamefont {H.}~\bibnamefont {Wilkens}}, \bibinfo {author}
  {\bibfnamefont {S.}~\bibnamefont {Bhandarkar}}, \bibinfo {author}
  {\bibfnamefont {A.}~\bibnamefont {Forsman}}, \bibinfo {author} {\bibfnamefont
  {S.}~\bibnamefont {Haan}}, \bibinfo {author} {\bibfnamefont {M.}~\bibnamefont
  {Hoppe}}, \bibinfo {author} {\bibfnamefont {H.}~\bibnamefont {Huang}},
  \bibinfo {author} {\bibfnamefont {E.}~\bibnamefont {Mapoles}}, \bibinfo
  {author} {\bibfnamefont {J.}~\bibnamefont {Moody}}, \bibinfo {author}
  {\bibfnamefont {J.}~\bibnamefont {Sater}}, \bibinfo {author} {\bibfnamefont
  {R.}~\bibnamefont {Seugling}}, \bibinfo {author} {\bibfnamefont
  {R.}~\bibnamefont {Stephens}}, \bibinfo {author} {\bibfnamefont
  {M.}~\bibnamefont {Takagi}}, \ and\ \bibinfo {author} {\bibfnamefont
  {H.}~\bibnamefont {Xu}},\ }\href@noop {} {\bibfield  {journal} {\bibinfo
  {journal} {Laser Part. Beams}\ }\textbf {\bibinfo {volume} {26}},\ \bibinfo
  {pages} {479} (\bibinfo {year} {2008})}\BibitemShut {NoStop}%
\bibitem [{\citenamefont {Barriga-Carrasco}(2008)}]{BarrigaPP08}%
  \BibitemOpen
  \bibfield  {author} {\bibinfo {author} {\bibfnamefont {M.~D.}\ \bibnamefont
  {Barriga-Carrasco}},\ }\href@noop {} {\bibfield  {journal} {\bibinfo
  {journal} {Phys. Plasmas}\ }\textbf {\bibinfo {volume} {15}},\ \bibinfo
  {pages} {033103} (\bibinfo {year} {2008})}\BibitemShut {NoStop}%
\bibitem [{\citenamefont {Chabot}\ \emph {et~al.}(1995)\citenamefont {Chabot},
  \citenamefont {Gardes}, \citenamefont {Box}, \citenamefont {Kiener},
  \citenamefont {Deutsch}, \citenamefont {Maynard}, \citenamefont {V.},
  \citenamefont {C.}, \citenamefont {D.},\ and\ \citenamefont
  {K.}}]{ChabotPRE95}%
  \BibitemOpen
  \bibfield  {author} {\bibinfo {author} {\bibfnamefont {M.}~\bibnamefont
  {Chabot}}, \bibinfo {author} {\bibfnamefont {D.}~\bibnamefont {Gardes}},
  \bibinfo {author} {\bibfnamefont {P.}~\bibnamefont {Box}}, \bibinfo {author}
  {\bibfnamefont {J.}~\bibnamefont {Kiener}}, \bibinfo {author} {\bibfnamefont
  {C.}~\bibnamefont {Deutsch}}, \bibinfo {author} {\bibfnamefont
  {G.}~\bibnamefont {Maynard}}, \bibinfo {author} {\bibfnamefont
  {A.}~\bibnamefont {V.}}, \bibinfo {author} {\bibfnamefont {F.}~\bibnamefont
  {C.}}, \bibinfo {author} {\bibfnamefont {H.}~\bibnamefont {D.}}, \ and\
  \bibinfo {author} {\bibfnamefont {W.}~\bibnamefont {K.}},\ }\href@noop {}
  {\bibfield  {journal} {\bibinfo  {journal} {Phys. Rev. E}\ }\textbf {\bibinfo
  {volume} {51}},\ \bibinfo {pages} {3504} (\bibinfo {year}
  {1995})}\BibitemShut {NoStop}%
\bibitem [{\citenamefont {Oguri}\ \emph {et~al.}(2000)\citenamefont {Oguri},
  \citenamefont {Tsubuke}, \citenamefont {Sakumi}, \citenamefont {Shibata},
  \citenamefont {Sato}, \citenamefont {Nishigori}, \citenamefont {Hasegawa},\
  and\ \citenamefont {Ogawa}}]{OguriNIMB00}%
  \BibitemOpen
  \bibfield  {author} {\bibinfo {author} {\bibfnamefont {Y.}~\bibnamefont
  {Oguri}}, \bibinfo {author} {\bibfnamefont {K.}~\bibnamefont {Tsubuke}},
  \bibinfo {author} {\bibfnamefont {A.}~\bibnamefont {Sakumi}}, \bibinfo
  {author} {\bibfnamefont {K.}~\bibnamefont {Shibata}}, \bibinfo {author}
  {\bibfnamefont {R.}~\bibnamefont {Sato}}, \bibinfo {author} {\bibfnamefont
  {K.}~\bibnamefont {Nishigori}}, \bibinfo {author} {\bibfnamefont
  {J.}~\bibnamefont {Hasegawa}}, \ and\ \bibinfo {author} {\bibfnamefont
  {M.}~\bibnamefont {Ogawa}},\ }\href@noop {} {\bibfield  {journal} {\bibinfo
  {journal} {Nucl. Instrum. Meth. B}\ }\textbf {\bibinfo {volume} {161-163}},\
  \bibinfo {pages} {155} (\bibinfo {year} {2000})}\BibitemShut {NoStop}%
\bibitem [{\citenamefont {Weyrich}\ \emph {et~al.}(1989)\citenamefont
  {Weyrich}, \citenamefont {Hoffmann}, \citenamefont {Jacoby}, \citenamefont
  {Wahl}, \citenamefont {Noll}, \citenamefont {Haas}, \citenamefont {Kunze},
  \citenamefont {Bimbot}, \citenamefont {Gardes}, \citenamefont {Rivet},
  \citenamefont {Deutsch},\ and\ \citenamefont {Fleurier}}]{WeyrichNIMPRA89}%
  \BibitemOpen
  \bibfield  {author} {\bibinfo {author} {\bibfnamefont {K.}~\bibnamefont
  {Weyrich}}, \bibinfo {author} {\bibfnamefont {D.~H.~H.}\ \bibnamefont
  {Hoffmann}}, \bibinfo {author} {\bibfnamefont {J.}~\bibnamefont {Jacoby}},
  \bibinfo {author} {\bibfnamefont {H.}~\bibnamefont {Wahl}}, \bibinfo {author}
  {\bibfnamefont {R.}~\bibnamefont {Noll}}, \bibinfo {author} {\bibfnamefont
  {R.}~\bibnamefont {Haas}}, \bibinfo {author} {\bibfnamefont {H.}~\bibnamefont
  {Kunze}}, \bibinfo {author} {\bibfnamefont {R.}~\bibnamefont {Bimbot}},
  \bibinfo {author} {\bibfnamefont {D.}~\bibnamefont {Gardes}}, \bibinfo
  {author} {\bibfnamefont {M.~F.}\ \bibnamefont {Rivet}}, \bibinfo {author}
  {\bibfnamefont {C.}~\bibnamefont {Deutsch}}, \ and\ \bibinfo {author}
  {\bibfnamefont {C.}~\bibnamefont {Fleurier}},\ }\href@noop {} {\bibfield
  {journal} {\bibinfo  {journal} {Nucl. Instrum. Meth. A}\ }\textbf {\bibinfo
  {volume} {278}},\ \bibinfo {pages} {52} (\bibinfo {year} {1989})}\BibitemShut
  {NoStop}%
\bibitem [{\citenamefont {Hoffmann}\ \emph {et~al.}(1990)\citenamefont
  {Hoffmann}, \citenamefont {Weyrich}, \citenamefont {Wahl}, \citenamefont
  {Gardes}, \citenamefont {Bimbot},\ and\ \citenamefont
  {Fleurier}}]{HoffmannPRA90}%
  \BibitemOpen
  \bibfield  {author} {\bibinfo {author} {\bibfnamefont {D.~H.~H.}\
  \bibnamefont {Hoffmann}}, \bibinfo {author} {\bibfnamefont {K.}~\bibnamefont
  {Weyrich}}, \bibinfo {author} {\bibfnamefont {H.}~\bibnamefont {Wahl}},
  \bibinfo {author} {\bibfnamefont {D.}~\bibnamefont {Gardes}}, \bibinfo
  {author} {\bibfnamefont {R.}~\bibnamefont {Bimbot}}, \ and\ \bibinfo {author}
  {\bibfnamefont {C.}~\bibnamefont {Fleurier}},\ }\href@noop {} {\bibfield
  {journal} {\bibinfo  {journal} {Phys.\ Rev. A}\ }\textbf {\bibinfo {volume}
  {42}},\ \bibinfo {pages} {2313} (\bibinfo {year} {1990})}\BibitemShut
  {NoStop}%
\bibitem [{\citenamefont {Gardes}\ \emph
  {et~al.}(1992{\natexlab{a}})\citenamefont {Gardes}, \citenamefont
  {Servajean}, \citenamefont {Jubica}, \citenamefont {Fleurier}, \citenamefont
  {Hong}, \citenamefont {Deutsch},\ and\ \citenamefont
  {Maynard}}]{GardesPRA92}%
  \BibitemOpen
  \bibfield  {author} {\bibinfo {author} {\bibfnamefont {D.}~\bibnamefont
  {Gardes}}, \bibinfo {author} {\bibfnamefont {A.}~\bibnamefont {Servajean}},
  \bibinfo {author} {\bibfnamefont {B.}~\bibnamefont {Jubica}}, \bibinfo
  {author} {\bibfnamefont {C.}~\bibnamefont {Fleurier}}, \bibinfo {author}
  {\bibfnamefont {D.}~\bibnamefont {Hong}}, \bibinfo {author} {\bibfnamefont
  {C.}~\bibnamefont {Deutsch}}, \ and\ \bibinfo {author} {\bibfnamefont
  {G.}~\bibnamefont {Maynard}},\ }\href@noop {} {\bibfield  {journal} {\bibinfo
   {journal} {Phys.\ Rev. A}\ }\textbf {\bibinfo {volume} {46}},\ \bibinfo
  {pages} {5101} (\bibinfo {year} {1992}{\natexlab{a}})}\BibitemShut {NoStop}%
\bibitem [{\citenamefont {Dietrich}\ \emph {et~al.}(1992)\citenamefont
  {Dietrich}, \citenamefont {Hoffmann}, \citenamefont {Boggasch}, \citenamefont
  {Jacoby}, \citenamefont {Wahl}, \citenamefont {Elfers}, \citenamefont {Haas},
  \citenamefont {Dubenkov},\ and\ \citenamefont {Golubev}}]{DietrichPRL92}%
  \BibitemOpen
  \bibfield  {author} {\bibinfo {author} {\bibfnamefont {K.~G.}\ \bibnamefont
  {Dietrich}}, \bibinfo {author} {\bibfnamefont {D.~H.~H.}\ \bibnamefont
  {Hoffmann}}, \bibinfo {author} {\bibfnamefont {E.}~\bibnamefont {Boggasch}},
  \bibinfo {author} {\bibfnamefont {J.}~\bibnamefont {Jacoby}}, \bibinfo
  {author} {\bibfnamefont {H.}~\bibnamefont {Wahl}}, \bibinfo {author}
  {\bibfnamefont {M.}~\bibnamefont {Elfers}}, \bibinfo {author} {\bibfnamefont
  {C.~R.}\ \bibnamefont {Haas}}, \bibinfo {author} {\bibfnamefont {V.~P.}\
  \bibnamefont {Dubenkov}}, \ and\ \bibinfo {author} {\bibfnamefont {A.~A.}\
  \bibnamefont {Golubev}},\ }\href@noop {} {\bibfield  {journal} {\bibinfo
  {journal} {Phys.\ Rev. Lett.}\ }\textbf {\bibinfo {volume} {69}},\ \bibinfo
  {pages} {3623} (\bibinfo {year} {1992})}\BibitemShut {NoStop}%
\bibitem [{\citenamefont {Gardes}\ \emph
  {et~al.}(1992{\natexlab{b}})\citenamefont {Gardes}, \citenamefont {Bimbot},
  \citenamefont {Rivet}, \citenamefont {Servajean}, \citenamefont {Fleurier},
  \citenamefont {Hong}, \citenamefont {Deutsch},\ and\ \citenamefont
  {Maynard}}]{GardesPA92}%
  \BibitemOpen
  \bibfield  {author} {\bibinfo {author} {\bibfnamefont {D.}~\bibnamefont
  {Gardes}}, \bibinfo {author} {\bibfnamefont {R.}~\bibnamefont {Bimbot}},
  \bibinfo {author} {\bibfnamefont {M.~F.}\ \bibnamefont {Rivet}}, \bibinfo
  {author} {\bibfnamefont {A.}~\bibnamefont {Servajean}}, \bibinfo {author}
  {\bibfnamefont {C.}~\bibnamefont {Fleurier}}, \bibinfo {author}
  {\bibfnamefont {D.}~\bibnamefont {Hong}}, \bibinfo {author} {\bibfnamefont
  {C.}~\bibnamefont {Deutsch}}, \ and\ \bibinfo {author} {\bibfnamefont
  {G.}~\bibnamefont {Maynard}},\ }\href@noop {} {\bibfield  {journal} {\bibinfo
   {journal} {Part. Accel.}\ }\textbf {\bibinfo {volume} {37-38}},\ \bibinfo
  {pages} {361} (\bibinfo {year} {1992}{\natexlab{b}})}\BibitemShut {NoStop}%
\bibitem [{\citenamefont {Couillaud}\ \emph {et~al.}(1994)\citenamefont
  {Couillaud}, \citenamefont {Deicas}, \citenamefont {Nardin}, \citenamefont
  {Beuve}, \citenamefont {Guihaum\'e}, \citenamefont {Renaud}, \citenamefont
  {Cukier}, \citenamefont {Deutsch},\ and\ \citenamefont
  {Maynard}}]{CouillaudPRE94}%
  \BibitemOpen
  \bibfield  {author} {\bibinfo {author} {\bibfnamefont {C.}~\bibnamefont
  {Couillaud}}, \bibinfo {author} {\bibfnamefont {R.}~\bibnamefont {Deicas}},
  \bibinfo {author} {\bibfnamefont {P.}~\bibnamefont {Nardin}}, \bibinfo
  {author} {\bibfnamefont {M.~A.}\ \bibnamefont {Beuve}}, \bibinfo {author}
  {\bibfnamefont {J.~M.}\ \bibnamefont {Guihaum\'e}}, \bibinfo {author}
  {\bibfnamefont {R.}~\bibnamefont {Renaud}}, \bibinfo {author} {\bibfnamefont
  {M.}~\bibnamefont {Cukier}}, \bibinfo {author} {\bibfnamefont
  {C.}~\bibnamefont {Deutsch}}, \ and\ \bibinfo {author} {\bibfnamefont
  {G.}~\bibnamefont {Maynard}},\ }\href@noop {} {\bibfield  {journal} {\bibinfo
   {journal} {Phys.\ Rev. E}\ }\textbf {\bibinfo {volume} {49}},\ \bibinfo
  {pages} {1545} (\bibinfo {year} {1994})}\BibitemShut {NoStop}%
\bibitem [{\citenamefont {Jacoby}\ \emph {et~al.}(1995)\citenamefont {Jacoby},
  \citenamefont {Hoffmann}, \citenamefont {Laux}, \citenamefont {Muller},
  \citenamefont {Wahl}, \citenamefont {Weyrich}, \citenamefont {Boggasch},
  \citenamefont {Heimrich}, \citenamefont {Stockl}, \citenamefont {Wetzler},\
  and\ \citenamefont {Miyamoto}}]{JacobyPRL95}%
  \BibitemOpen
  \bibfield  {author} {\bibinfo {author} {\bibfnamefont {J.}~\bibnamefont
  {Jacoby}}, \bibinfo {author} {\bibfnamefont {D.~H.~H.}\ \bibnamefont
  {Hoffmann}}, \bibinfo {author} {\bibfnamefont {W.}~\bibnamefont {Laux}},
  \bibinfo {author} {\bibfnamefont {R.~W.}\ \bibnamefont {Muller}}, \bibinfo
  {author} {\bibfnamefont {H.}~\bibnamefont {Wahl}}, \bibinfo {author}
  {\bibfnamefont {K.}~\bibnamefont {Weyrich}}, \bibinfo {author} {\bibfnamefont
  {E.}~\bibnamefont {Boggasch}}, \bibinfo {author} {\bibfnamefont
  {B.}~\bibnamefont {Heimrich}}, \bibinfo {author} {\bibfnamefont
  {C.}~\bibnamefont {Stockl}}, \bibinfo {author} {\bibfnamefont
  {H.}~\bibnamefont {Wetzler}}, \ and\ \bibinfo {author} {\bibfnamefont
  {S.}~\bibnamefont {Miyamoto}},\ }\href@noop {} {\bibfield  {journal}
  {\bibinfo  {journal} {Phys. Rev. Lett.}\ }\textbf {\bibinfo {volume} {74}},\
  \bibinfo {pages} {1550} (\bibinfo {year} {1995})}\BibitemShut {NoStop}%
\bibitem [{\citenamefont {Kojima}\ \emph {et~al.}(2002)\citenamefont {Kojima},
  \citenamefont {Mitomo}, \citenamefont {Sasaki}, \citenamefont {Hasegawa},\
  and\ \citenamefont {Ogawa}}]{KojimaLPB02}%
  \BibitemOpen
  \bibfield  {author} {\bibinfo {author} {\bibfnamefont {M.}~\bibnamefont
  {Kojima}}, \bibinfo {author} {\bibfnamefont {M.}~\bibnamefont {Mitomo}},
  \bibinfo {author} {\bibfnamefont {T.}~\bibnamefont {Sasaki}}, \bibinfo
  {author} {\bibfnamefont {J.}~\bibnamefont {Hasegawa}}, \ and\ \bibinfo
  {author} {\bibfnamefont {M.}~\bibnamefont {Ogawa}},\ }\href@noop {}
  {\bibfield  {journal} {\bibinfo  {journal} {Laser Part. Beams}\ }\textbf
  {\bibinfo {volume} {20}},\ \bibinfo {pages} {475} (\bibinfo {year}
  {2002})}\BibitemShut {NoStop}%
\bibitem [{\citenamefont {Skobelev}\ \emph {et~al.}(2005)\citenamefont
  {Skobelev}, \citenamefont {Kalpakchieva}, \citenamefont {Astabatyan},
  \citenamefont {Vincour}, \citenamefont {Kulko}, \citenamefont {Lobastov},
  \citenamefont {Lukyanov}, \citenamefont {Markaryan}, \citenamefont {Maslov},
  \citenamefont {Sobolev},\ and\ \citenamefont {Ugryumov}}]{SkobelevNIMB05}%
  \BibitemOpen
  \bibfield  {author} {\bibinfo {author} {\bibfnamefont {N.~K.}\ \bibnamefont
  {Skobelev}}, \bibinfo {author} {\bibfnamefont {R.}~\bibnamefont
  {Kalpakchieva}}, \bibinfo {author} {\bibfnamefont {R.~A.}\ \bibnamefont
  {Astabatyan}}, \bibinfo {author} {\bibfnamefont {J.}~\bibnamefont {Vincour}},
  \bibinfo {author} {\bibfnamefont {A.~A.}\ \bibnamefont {Kulko}}, \bibinfo
  {author} {\bibfnamefont {S.~P.}\ \bibnamefont {Lobastov}}, \bibinfo {author}
  {\bibfnamefont {S.~M.}\ \bibnamefont {Lukyanov}}, \bibinfo {author}
  {\bibfnamefont {E.~R.}\ \bibnamefont {Markaryan}}, \bibinfo {author}
  {\bibfnamefont {V.~A.}\ \bibnamefont {Maslov}}, \bibinfo {author}
  {\bibfnamefont {Y.~H.}\ \bibnamefont {Sobolev}}, \ and\ \bibinfo {author}
  {\bibfnamefont {V.~Y.}\ \bibnamefont {Ugryumov}},\ }\href@noop {} {\bibfield
  {journal} {\bibinfo  {journal} {Nucl. Instrum. Meth. B}\ }\textbf {\bibinfo
  {volume} {227}},\ \bibinfo {pages} {471} (\bibinfo {year}
  {2005})}\BibitemShut {NoStop}%
\bibitem [{\citenamefont {Frank}\ \emph {et~al.}(2013)\citenamefont {Frank},
  \citenamefont {Blazevi\'c}, \citenamefont {Bagnoud}, \citenamefont {Basko},
  \citenamefont {Borner}, \citenamefont {Cayzac}, \citenamefont {Kraus},
  \citenamefont {Hessling}, \citenamefont {Hoffmann}, \citenamefont {Ortner},
  \citenamefont {Otten}, \citenamefont {Pelka}, \citenamefont {Pepler},
  \citenamefont {Schumacher}, \citenamefont {Tauschwitz},\ and\ \citenamefont
  {Roth}}]{FrankPRL13}%
  \BibitemOpen
  \bibfield  {author} {\bibinfo {author} {\bibfnamefont {A.}~\bibnamefont
  {Frank}}, \bibinfo {author} {\bibfnamefont {A.}~\bibnamefont {Blazevi\'c}},
  \bibinfo {author} {\bibfnamefont {V.}~\bibnamefont {Bagnoud}}, \bibinfo
  {author} {\bibfnamefont {M.~M.}\ \bibnamefont {Basko}}, \bibinfo {author}
  {\bibfnamefont {M.}~\bibnamefont {Borner}}, \bibinfo {author} {\bibfnamefont
  {W.}~\bibnamefont {Cayzac}}, \bibinfo {author} {\bibfnamefont
  {D.}~\bibnamefont {Kraus}}, \bibinfo {author} {\bibfnamefont
  {T.}~\bibnamefont {Hessling}}, \bibinfo {author} {\bibfnamefont {D.~H.~H.}\
  \bibnamefont {Hoffmann}}, \bibinfo {author} {\bibfnamefont {A.}~\bibnamefont
  {Ortner}}, \bibinfo {author} {\bibfnamefont {A.}~\bibnamefont {Otten}},
  \bibinfo {author} {\bibfnamefont {A.}~\bibnamefont {Pelka}}, \bibinfo
  {author} {\bibfnamefont {D.}~\bibnamefont {Pepler}}, \bibinfo {author}
  {\bibfnamefont {D.}~\bibnamefont {Schumacher}}, \bibinfo {author}
  {\bibfnamefont {A.}~\bibnamefont {Tauschwitz}}, \ and\ \bibinfo {author}
  {\bibfnamefont {M.}~\bibnamefont {Roth}},\ }\href@noop {} {\bibfield
  {journal} {\bibinfo  {journal} {Phys.\ Rev. Lett.}\ }\textbf {\bibinfo
  {volume} {110}},\ \bibinfo {pages} {115001} (\bibinfo {year}
  {2013})}\BibitemShut {NoStop}%
\bibitem [{\citenamefont {Gauthier}\ \emph {et~al.}(2013)\citenamefont
  {Gauthier}, \citenamefont {Chen}, \citenamefont {Levy}, \citenamefont
  {Audebert}, \citenamefont {Blancard}, \citenamefont {Ceccotti}, \citenamefont
  {Cerchez}, \citenamefont {Doria}, \citenamefont {Floquet}, \citenamefont
  {Lamour}, \citenamefont {Peth}, \citenamefont {Romagnani}, \citenamefont
  {Rozet}, \citenamefont {Scheinder}, \citenamefont {Shepherd}, \citenamefont
  {Toncian}, \citenamefont {Vernhet}, \citenamefont {Willi}, \citenamefont
  {Borghesi}, \citenamefont {Faussurier},\ and\ \citenamefont
  {Fuchs}}]{GauthierPRL13}%
  \BibitemOpen
  \bibfield  {author} {\bibinfo {author} {\bibfnamefont {M.}~\bibnamefont
  {Gauthier}}, \bibinfo {author} {\bibfnamefont {S.~N.}\ \bibnamefont {Chen}},
  \bibinfo {author} {\bibfnamefont {A.}~\bibnamefont {Levy}}, \bibinfo {author}
  {\bibfnamefont {P.}~\bibnamefont {Audebert}}, \bibinfo {author}
  {\bibfnamefont {C.}~\bibnamefont {Blancard}}, \bibinfo {author}
  {\bibfnamefont {T.}~\bibnamefont {Ceccotti}}, \bibinfo {author}
  {\bibfnamefont {M.}~\bibnamefont {Cerchez}}, \bibinfo {author} {\bibfnamefont
  {D.}~\bibnamefont {Doria}}, \bibinfo {author} {\bibfnamefont
  {V.}~\bibnamefont {Floquet}}, \bibinfo {author} {\bibfnamefont
  {E.}~\bibnamefont {Lamour}}, \bibinfo {author} {\bibfnamefont
  {C.}~\bibnamefont {Peth}}, \bibinfo {author} {\bibfnamefont {L.}~\bibnamefont
  {Romagnani}}, \bibinfo {author} {\bibfnamefont {J.~P.}\ \bibnamefont
  {Rozet}}, \bibinfo {author} {\bibfnamefont {M.}~\bibnamefont {Scheinder}},
  \bibinfo {author} {\bibfnamefont {R.}~\bibnamefont {Shepherd}}, \bibinfo
  {author} {\bibfnamefont {T.}~\bibnamefont {Toncian}}, \bibinfo {author}
  {\bibfnamefont {D.}~\bibnamefont {Vernhet}}, \bibinfo {author} {\bibfnamefont
  {O.}~\bibnamefont {Willi}}, \bibinfo {author} {\bibfnamefont
  {M.}~\bibnamefont {Borghesi}}, \bibinfo {author} {\bibfnamefont
  {G.}~\bibnamefont {Faussurier}}, \ and\ \bibinfo {author} {\bibfnamefont
  {J.}~\bibnamefont {Fuchs}},\ }\href@noop {} {\bibfield  {journal} {\bibinfo
  {journal} {Phys. Rev. Lett.}\ }\textbf {\bibinfo {volume} {110}},\ \bibinfo
  {pages} {135003} (\bibinfo {year} {2013})}\BibitemShut {NoStop}%
\bibitem [{\citenamefont {Nardi}\ and\ \citenamefont
  {Zinamon}(1982)}]{NardiPRL82}%
  \BibitemOpen
  \bibfield  {author} {\bibinfo {author} {\bibfnamefont {E.}~\bibnamefont
  {Nardi}}\ and\ \bibinfo {author} {\bibfnamefont {Z.}~\bibnamefont
  {Zinamon}},\ }\href@noop {} {\bibfield  {journal} {\bibinfo  {journal} {Phys.
  Rev. Lett.}\ }\textbf {\bibinfo {volume} {49}},\ \bibinfo {pages} {1251}
  (\bibinfo {year} {1982})}\BibitemShut {NoStop}%
\bibitem [{\citenamefont {Peter}\ and\ \citenamefont
  {{Meyer-ter-Vehn}}(1991)}]{PeterPRA91b}%
  \BibitemOpen
  \bibfield  {author} {\bibinfo {author} {\bibfnamefont {T.}~\bibnamefont
  {Peter}}\ and\ \bibinfo {author} {\bibfnamefont {J.}~\bibnamefont
  {{Meyer-ter-Vehn}}},\ }\href@noop {} {\bibfield  {journal} {\bibinfo
  {journal} {Phys. Rev. A}\ }\textbf {\bibinfo {volume} {43}},\ \bibinfo
  {pages} {2015} (\bibinfo {year} {1991})}\BibitemShut {NoStop}%
\bibitem [{\citenamefont {Faussurier}\ \emph {et~al.}(1997)\citenamefont
  {Faussurier}, \citenamefont {Blancard},\ and\ \citenamefont
  {Decoster}}]{FaussurierJQSRT97}%
  \BibitemOpen
  \bibfield  {author} {\bibinfo {author} {\bibfnamefont {G.}~\bibnamefont
  {Faussurier}}, \bibinfo {author} {\bibfnamefont {C.}~\bibnamefont
  {Blancard}}, \ and\ \bibinfo {author} {\bibfnamefont {A.}~\bibnamefont
  {Decoster}},\ }\href@noop {} {\bibfield  {journal} {\bibinfo  {journal} {J.
  Quant. Spectrosc. Radiat. Transfer}\ }\textbf {\bibinfo {volume} {58}},\
  \bibinfo {pages} {233} (\bibinfo {year} {1997})}\BibitemShut {NoStop}%
\bibitem [{\citenamefont {Slater}(1930)}]{SlaterPR30}%
  \BibitemOpen
  \bibfield  {author} {\bibinfo {author} {\bibfnamefont {J.~C.}\ \bibnamefont
  {Slater}},\ }\href@noop {} {\bibfield  {journal} {\bibinfo  {journal} {Phys.
  Rev.}\ }\textbf {\bibinfo {volume} {36}},\ \bibinfo {pages} {57} (\bibinfo
  {year} {1930})}\BibitemShut {NoStop}%
\bibitem [{\citenamefont {Gryzinski}(1965{\natexlab{a}})}]{GryzinskiPR65a}%
  \BibitemOpen
  \bibfield  {author} {\bibinfo {author} {\bibfnamefont {M.}~\bibnamefont
  {Gryzinski}},\ }\href@noop {} {\bibfield  {journal} {\bibinfo  {journal}
  {Phys. Rev.}\ }\textbf {\bibinfo {volume} {138}},\ \bibinfo {pages} {A336}
  (\bibinfo {year} {1965}{\natexlab{a}})}\BibitemShut {NoStop}%
\bibitem [{\citenamefont {Gryzinski}(1965{\natexlab{b}})}]{GryzinskiPR65b}%
  \BibitemOpen
  \bibfield  {author} {\bibinfo {author} {\bibfnamefont {M.}~\bibnamefont
  {Gryzinski}},\ }\href@noop {} {\bibfield  {journal} {\bibinfo  {journal}
  {Phys. Rev.}\ }\textbf {\bibinfo {volume} {138}},\ \bibinfo {pages} {A322}
  (\bibinfo {year} {1965}{\natexlab{b}})}\BibitemShut {NoStop}%
\bibitem [{\citenamefont {McGuire}\ and\ \citenamefont
  {Richard}(1973)}]{McGuirePRA73}%
  \BibitemOpen
  \bibfield  {author} {\bibinfo {author} {\bibfnamefont {J.~H.}\ \bibnamefont
  {McGuire}}\ and\ \bibinfo {author} {\bibfnamefont {P.}~\bibnamefont
  {Richard}},\ }\href@noop {} {\bibfield  {journal} {\bibinfo  {journal} {Phys.
  Rev. A}\ }\textbf {\bibinfo {volume} {8}},\ \bibinfo {pages} {1374} (\bibinfo
  {year} {1973})}\BibitemShut {NoStop}%
\bibitem [{\citenamefont {Lotz}(1967)}]{LotzZP67}%
  \BibitemOpen
  \bibfield  {author} {\bibinfo {author} {\bibfnamefont {W.}~\bibnamefont
  {Lotz}},\ }\href@noop {} {\bibfield  {journal} {\bibinfo  {journal} {Z.
  Phys.}\ }\textbf {\bibinfo {volume} {206}},\ \bibinfo {pages} {205} (\bibinfo
  {year} {1967})}\BibitemShut {NoStop}%
\bibitem [{\citenamefont {Lotz}(1968)}]{LotzZP68}%
  \BibitemOpen
  \bibfield  {author} {\bibinfo {author} {\bibfnamefont {W.}~\bibnamefont
  {Lotz}},\ }\href@noop {} {\bibfield  {journal} {\bibinfo  {journal} {Z.
  Phys.}\ }\textbf {\bibinfo {volume} {216}},\ \bibinfo {pages} {241} (\bibinfo
  {year} {1968})}\BibitemShut {NoStop}%
\bibitem [{\citenamefont {Menzel}(1937)}]{MenzelAJ37}%
  \BibitemOpen
  \bibfield  {author} {\bibinfo {author} {\bibfnamefont {D.~H.}\ \bibnamefont
  {Menzel}},\ }\href@noop {} {\bibfield  {journal} {\bibinfo  {journal}
  {Astrophys. J.}\ }\textbf {\bibinfo {volume} {85}},\ \bibinfo {pages} {330}
  (\bibinfo {year} {1937})}\BibitemShut {NoStop}%
\bibitem [{\citenamefont {Jr.}(1948)}]{SpitzerAJ48}%
  \BibitemOpen
  \bibfield  {author} {\bibinfo {author} {\bibfnamefont {L.~S.}\ \bibnamefont
  {Jr.}},\ }\href@noop {} {\bibfield  {journal} {\bibinfo  {journal}
  {Astrophys. J.}\ }\textbf {\bibinfo {volume} {107}},\ \bibinfo {pages} {7}
  (\bibinfo {year} {1948})}\BibitemShut {NoStop}%
\bibitem [{\citenamefont {Seaton}(1959)}]{SeatonRAS59}%
  \BibitemOpen
  \bibfield  {author} {\bibinfo {author} {\bibfnamefont {M.~J.}\ \bibnamefont
  {Seaton}},\ }\href@noop {} {\bibfield  {journal} {\bibinfo  {journal} {Mon.
  Not. R. Astron. Soc.}\ }\textbf {\bibinfo {volume} {119}},\ \bibinfo {pages}
  {81} (\bibinfo {year} {1959})}\BibitemShut {NoStop}%
\bibitem [{\citenamefont {Peter}(1990)}]{PeterLPB90}%
  \BibitemOpen
  \bibfield  {author} {\bibinfo {author} {\bibfnamefont {T.}~\bibnamefont
  {Peter}},\ }\href@noop {} {\bibfield  {journal} {\bibinfo  {journal} {Laser
  Part. Beams}\ }\textbf {\bibinfo {volume} {8}},\ \bibinfo {pages} {643}
  (\bibinfo {year} {1990})}\BibitemShut {NoStop}%
\bibitem [{\citenamefont {Hahn}(1980)}]{HahnPRA80}%
  \BibitemOpen
  \bibfield  {author} {\bibinfo {author} {\bibfnamefont {Y.}~\bibnamefont
  {Hahn}},\ }\href@noop {} {\bibfield  {journal} {\bibinfo  {journal} {Phys.
  Rev. A.}\ }\textbf {\bibinfo {volume} {22}},\ \bibinfo {pages} {2896}
  (\bibinfo {year} {1980})}\BibitemShut {NoStop}%
\bibitem [{\citenamefont {Peter}\ \emph {et~al.}(1986)\citenamefont {Peter},
  \citenamefont {Arnold},\ and\ \citenamefont {{Meyer-ter-Vehn}}}]{PeterPRL86}%
  \BibitemOpen
  \bibfield  {author} {\bibinfo {author} {\bibfnamefont {T.}~\bibnamefont
  {Peter}}, \bibinfo {author} {\bibfnamefont {R.}~\bibnamefont {Arnold}}, \
  and\ \bibinfo {author} {\bibfnamefont {J.}~\bibnamefont {{Meyer-ter-Vehn}}},\
  }\href@noop {} {\bibfield  {journal} {\bibinfo  {journal} {Phys. Rev. Lett.}\
  }\textbf {\bibinfo {volume} {57}},\ \bibinfo {pages} {1859} (\bibinfo {year}
  {1986})}\BibitemShut {NoStop}%
\bibitem [{\citenamefont {Dasgupta}\ and\ \citenamefont
  {Whitney}(1990)}]{DasguptaPRA90}%
  \BibitemOpen
  \bibfield  {author} {\bibinfo {author} {\bibfnamefont {A.}~\bibnamefont
  {Dasgupta}}\ and\ \bibinfo {author} {\bibfnamefont {K.~G.}\ \bibnamefont
  {Whitney}},\ }\href@noop {} {\bibfield  {journal} {\bibinfo  {journal} {Phys.
  Rev. A}\ }\textbf {\bibinfo {volume} {42}},\ \bibinfo {pages} {2640}
  (\bibinfo {year} {1990})}\BibitemShut {NoStop}%
\bibitem [{\citenamefont {Fournier}\ \emph {et~al.}(1997)\citenamefont
  {Fournier}, \citenamefont {Cohen},\ and\ \citenamefont
  {Goldstein}}]{FournierPRA97}%
  \BibitemOpen
  \bibfield  {author} {\bibinfo {author} {\bibfnamefont {K.~B.}\ \bibnamefont
  {Fournier}}, \bibinfo {author} {\bibfnamefont {M.}~\bibnamefont {Cohen}}, \
  and\ \bibinfo {author} {\bibfnamefont {W.~H.}\ \bibnamefont {Goldstein}},\
  }\href@noop {} {\bibfield  {journal} {\bibinfo  {journal} {Phys. Rev. A}\
  }\textbf {\bibinfo {volume} {56}},\ \bibinfo {pages} {4715} (\bibinfo {year}
  {1997})}\BibitemShut {NoStop}%
\bibitem [{\citenamefont {Bethe}\ and\ \citenamefont
  {Salpeter}(1957)}]{QMOTEA_Bethe}%
  \BibitemOpen
  \bibfield  {author} {\bibinfo {author} {\bibfnamefont {H.~A.}\ \bibnamefont
  {Bethe}}\ and\ \bibinfo {author} {\bibfnamefont {E.~E.}\ \bibnamefont
  {Salpeter}},\ }\href@noop {} {\emph {\bibinfo {title} {Quantum mechanics of
  one- and two-electron atoms}}}\ (\bibinfo  {publisher} {Academic Press Inc,
  New York},\ \bibinfo {year} {1957})\BibitemShut {NoStop}%
\bibitem [{\citenamefont {Zel'dovich}\ and\ \citenamefont
  {Raizer}(1966)}]{ZeldovichRaizer66}%
  \BibitemOpen
  \bibfield  {author} {\bibinfo {author} {\bibfnamefont {Y.~B.}\ \bibnamefont
  {Zel'dovich}}\ and\ \bibinfo {author} {\bibfnamefont {Y.}~\bibnamefont
  {Raizer}},\ }\enquote {\bibinfo {title} {Physics of shock waves and
  high-temperature hydrodynamic phenomena},}\ \ (\bibinfo  {publisher}
  {Academic, New York},\ \bibinfo {year} {1966})\ p.\ \bibinfo {pages}
  {406}\BibitemShut {NoStop}%
\bibitem [{\citenamefont {Massey}\ and\ \citenamefont
  {Burhop}(1952)}]{Massey52}%
  \BibitemOpen
  \bibfield  {author} {\bibinfo {author} {\bibfnamefont {H.~S.~W.}\
  \bibnamefont {Massey}}\ and\ \bibinfo {author} {\bibfnamefont {E.~H.~S.}\
  \bibnamefont {Burhop}},\ }\href@noop {} {\emph {\bibinfo {title} {Electronic
  and Ionic Impact Phenomena}}}\ (\bibinfo  {publisher} {Oxford University
  Press, London},\ \bibinfo {year} {1952})\BibitemShut {NoStop}%
\bibitem [{\citenamefont {Bohr}(1940)}]{Bohr40}%
  \BibitemOpen
  \bibfield  {author} {\bibinfo {author} {\bibfnamefont {N.}~\bibnamefont
  {Bohr}},\ }\href@noop {} {\bibfield  {journal} {\bibinfo  {journal} {Phys.
  Rev.}\ }\textbf {\bibinfo {volume} {58}},\ \bibinfo {pages} {654} (\bibinfo
  {year} {1940})}\BibitemShut {NoStop}%
\bibitem [{\citenamefont {Bohr}(1941)}]{Bohr41}%
  \BibitemOpen
  \bibfield  {author} {\bibinfo {author} {\bibfnamefont {N.}~\bibnamefont
  {Bohr}},\ }\href@noop {} {\bibfield  {journal} {\bibinfo  {journal} {Phys.
  Rev.}\ }\textbf {\bibinfo {volume} {59}},\ \bibinfo {pages} {270} (\bibinfo
  {year} {1941})}\BibitemShut {NoStop}%
\bibitem [{\citenamefont {Lamb}(1940)}]{Lamb40}%
  \BibitemOpen
  \bibfield  {author} {\bibinfo {author} {\bibfnamefont {W.~E.}\ \bibnamefont
  {Lamb}},\ }\href@noop {} {\bibfield  {journal} {\bibinfo  {journal} {Phys.
  Rev.}\ }\textbf {\bibinfo {volume} {58}},\ \bibinfo {pages} {696} (\bibinfo
  {year} {1940})}\BibitemShut {NoStop}%
\bibitem [{\citenamefont {Takamoto}\ and\ \citenamefont
  {Kaneko}(1999)}]{TakamotoNIMB99}%
  \BibitemOpen
  \bibfield  {author} {\bibinfo {author} {\bibfnamefont {T.}~\bibnamefont
  {Takamoto}}\ and\ \bibinfo {author} {\bibfnamefont {T.}~\bibnamefont
  {Kaneko}},\ }\href@noop {} {\bibfield  {journal} {\bibinfo  {journal} {Nucl.
  Instum. Meth. B}\ }\textbf {\bibinfo {volume} {153}},\ \bibinfo {pages} {21}
  (\bibinfo {year} {1999})}\BibitemShut {NoStop}%
\bibitem [{\citenamefont {Barriga-Carrasco}(2010)}]{BarrigaPRE10}%
  \BibitemOpen
  \bibfield  {author} {\bibinfo {author} {\bibfnamefont {M.~D.}\ \bibnamefont
  {Barriga-Carrasco}},\ }\href@noop {} {\bibfield  {journal} {\bibinfo
  {journal} {Phys.\ Rev. E}\ }\textbf {\bibinfo {volume} {82}},\ \bibinfo
  {pages} {046403} (\bibinfo {year} {2010})}\BibitemShut {NoStop}%
\bibitem [{\citenamefont {Barriga-Carrasco}\ \emph {et~al.}(2016)\citenamefont
  {Barriga-Carrasco}, \citenamefont {Casas},\ and\ \citenamefont
  {Morales}}]{BarrigaPRE16}%
  \BibitemOpen
  \bibfield  {author} {\bibinfo {author} {\bibfnamefont {M.~D.}\ \bibnamefont
  {Barriga-Carrasco}}, \bibinfo {author} {\bibfnamefont {D.}~\bibnamefont
  {Casas}}, \ and\ \bibinfo {author} {\bibfnamefont {R.}~\bibnamefont
  {Morales}},\ }\href@noop {} {\bibfield  {journal} {\bibinfo  {journal}
  {Phys.\ Rev. E}\ }\textbf {\bibinfo {volume} {93}},\ \bibinfo {pages}
  {033204} (\bibinfo {year} {2016})}\BibitemShut {NoStop}%
\bibitem [{\citenamefont {Brandt}\ and\ \citenamefont
  {Kitagawa}(1982)}]{BrandtPRB82}%
  \BibitemOpen
  \bibfield  {author} {\bibinfo {author} {\bibfnamefont {W.}~\bibnamefont
  {Brandt}}\ and\ \bibinfo {author} {\bibfnamefont {M.}~\bibnamefont
  {Kitagawa}},\ }\href@noop {} {\bibfield  {journal} {\bibinfo  {journal}
  {Phys. Rev. B}\ }\textbf {\bibinfo {volume} {25}},\ \bibinfo {pages} {5631}
  (\bibinfo {year} {1982})}\BibitemShut {NoStop}%
\end{thebibliography}%

\end{document}